\documentclass[structabstract]{aa}  
\usepackage{graphicx}
\usepackage{textcomp}
\usepackage{longtable,lscape}
\usepackage[varg]{txfonts}
\usepackage{mathrsfs}
\usepackage{natbib}
\usepackage[colorlinks, citecolor={blue}]{hyperref}
\titlerunning{Effects of the environment on galaxies in the CIG: physical satellites and large scale structure}

\begin{document}
   \title{Effects of the environment on galaxies in the Catalogue of Isolated Galaxies: physical satellites and large scale structure}

   \author{M.\,Argudo-Fern\'andez\inst{1,2}
          \and
          S.\,Verley\inst{2}
          \and
          G.\,Bergond\inst{3}
          \and
          J.\,Sulentic\inst{1}
          \and
          J.\,Sabater\inst{4}
          \and
          M.\,Fern\'andez\,Lorenzo\inst{1}
          \and
          D.\,Espada\inst{5,6,7}
          \and          
          S.\,Leon\inst{5}
          \and
           S.\,S\'anchez-Exp\'osito\inst{1}
          \and
          J.\,D.\,Santander-Vela\inst{1}           
          \and
           L.\,Verdes-Montenegro\inst{1}
          }

   \institute{Instituto de Astrof\'isica de Andaluc\'ia (CSIC) Apdo. 3004, 18080 Granada, Spain%\\
              %\email{margudo@iaa.es}
         \and
             Departamento de F\'isica Te\'orica y del Cosmos, Universidad de Granada, 18071 Granada, Spain%\\
             %\email{simon@ugr.es}        
         \and
            Centro Astron\'omico Hispano Alem\'an, Calar Alto, (CSIC-MPG), C/ Jes\'us Durb\'an Rem\'on 2-2, E-04004 Almer\'ia, Spain    
         \and
             Institute for Astronomy, University of Edinburgh, Edinburgh EH9 3HJ, UK
         \and
             Joint ALMA Observatory (ALMA/ESO), Alonso de C\'ordova 3107, Vitacura, Santiago 763-0355, Chile 
         \and
             National Astronomical Observatory of Japan (NAOJ), 2-21-1 Osawa, Mitaka, Tokyo 181-8588, Japan
         \and      
             Department of Astronomical Science, The Graduate University for Advanced Studies (SOKENDAI), 2-21-1 Osawa, Mitaka, Tokyo 181-8588, Japan             
             }
            
   \date{Received August 16, 2013; accepted February 22, 2014}

% \abstract{}{}{}{}{} 
% 5 {} token are mandatory
 
\abstract
{We present a study of the 3-dimensional environment for a sample of 386 galaxies in the \textbf{C}atalogue of \textbf{I}solated \textbf{G}alaxies (CIG, Karachentseva 1973) using the Ninth Data Release of the Sloan Digital Sky Survey (SDSS-DR9).}
{We aim to identify and quantify the effects of the satellite distribution around a sample of galaxies in the CIG, as well as the effects of the Large Scale Structure (LSS).}
{To recover the physically bound galaxies we first focus on the satellites which are within the escape speed of each CIG galaxy. We also propose a more conservative method using the stacked Gaussian distribution of the velocity difference of the neighbours. The tidal strengths affecting the primary galaxy are estimated to quantify the effects of the local and LSS environments. We also define the projected number density parameter at the 5$^{\rm th}$ nearest neighbour to characterise the LSS around the CIG galaxies.}
{Out of the 386 CIG galaxies considered in this study, at least 340 (88\% of the sample) have no physically linked satellite.  
Following the more conservative Gaussian distribution of physical satellites around the CIG galaxies leads to upper limits. Out of the 386 CIG galaxies, 327 (85\% of the sample) have no physical companion within a projected distance of 0.3\,Mpc. The CIG galaxies are distributed following the LSS of the local Universe, although presenting a large heterogeneity in their degree of connection with it. When present around a CIG galaxy, the effect of physically bound galaxies largely dominates (usually by more than 90\%) the tidal strengths generated by the LSS.}
{The CIG samples a variety of environments, from galaxies with physical satellites to galaxies with no neighbours within 3\,Mpc.
A clear segregation appears between early-type CIG galaxies with companions and isolated late-type CIG galaxies. Isolated galaxies are in general bluer, with likely younger stellar populations and rather high star formation with respect to older, redder CIG galaxies with companions. Reciprocally, the satellites are redder and with an older stellar populations around massive early-type CIG galaxies, while they have a younger stellar content around massive late-type CIG galaxies. This suggests that the CIG is composed of a heterogeneous population of galaxies, sampling from old to more recent, dynamical systems of galaxies. CIG galaxies with companions might have a mild tendency (0.3-0.4 dex) to be more massive, and may indicate a higher frequency of having suffered a merger in the past.}
   

   \maketitle
%
%________________________________________________________________

\section{Introduction}  \label{Sec:intro}
   
Isolated galaxies are located, by definition, in low-density regions of the Universe, and should not be significantly influenced by their neighbours. Does a separate population of isolated galaxies exist, or are isolated galaxies simply the least clustered galaxies of the Large Scale Structure (LSS)? It is assumed that over the past several billion years the evolution of these objects has largely been driven by internal processes. A significant population of isolated galaxies is of great interest for testing different scenarios of the origin and evolution of galaxies. In this sense, isolated galaxies are an ideal sample of reference for studying the effects of environment on different galaxy properties. Such a sample would represent the most nurture-free galaxy population. 

Studies of isolated galaxies can be argued to begin with the publication of the Catalogue of Isolated Galaxies \citep[CIG;][]{1973AISAO...8....3K}. The AMIGA (\textbf{A}nalysis of the interstellar \textbf{M}edium of \textbf{I}solated \textbf{GA}laxies\footnote{\texttt{http://amiga.iaa.es}}) project \citep{2005A&A...436..443V} is based upon a re-evaluation of the CIG. It is a first step in trying to identify and better characterise isolated galaxies in the local Universe. \citet{2005A&A...436..443V} argued that 50\% or more galaxies in the CIG show a homogeneous redshift distribution. \citet{2006A&A...449..937S}, and more recently \citet{2012A&A...540A..47F}, found that 2/3 of the CIG are Sb-Sc late-type galaxies, and 14\% are early-type. This implies an extremely high late-type fraction and extremely low early-type population. At intermediate redshift, \citet{2012MNRAS.419.3018C} found that early-type systems in higher density regions tend to be more extended than their counterparts  in low density environments. Taking into account the effect of the local environment, \citet{2013MNRAS.434..325F} show that the number of satellites around a galaxy affects its size. CIG galaxies have larger sizes than galaxies in the \citet{2010ApJS..186..427N} sample with zero or one satellite, which are also larger than galaxies in \citet{2010ApJS..186..427N} with two or more satellites.

The distribution of satellites (faint companions) around isolated primary galaxies provides important information about galaxy formation, as well as a critical test of the $\Lambda$CDM model on small scales \citep{1987MNRAS.226..543E,2007AAS...21112602C,2010ApJ...709.1321A,2012MNRAS.425.2817F,2013JCAP...03..014A,2013ApJ...772..109B}. 
This explains the growing interest for studying the satellite distribution \citep{2003ApJ...598..260P,2005MNRAS.356.1045S,2012MNRAS.427..428G}, and for exploring the link between galaxy properties and the satellite population \citep{2007ApJ...658..898P,2011MNRAS.417..370G,2011AstBu..66..389K,2012MNRAS.424.1454E,2013ApJ...770...96G}.

According to a previous study \citep{2013A&A...560A...9A}, the criteria proposed by \citet{1973AISAO...8....3K} to remove fore- and background galaxies are not fully efficient. About 50\% of the neighbours, considered as potential companions, have very high recession velocities with respect to the central CIG galaxy: the condition is too restrictive, and may consider as not isolated galaxies slightly affected by their environment. On the other hand, about 92\% of neighbour galaxies showing recession velocities similar to the corresponding CIG galaxy are not considered as potential companions by the CIG isolation criteria, and may have a non negligible influence on the evolution of the central CIG galaxy. This motivates us to extend the study, taking into account nearby and similar redshift companions to identify physical satellites affecting the evolution of the central CIG galaxy, so as to provide a more physical estimation of the isolation degree of the CIG. About 60\% of the CIG galaxies have no major (similar-size) companion in the SDSS, according to the CIG (purely photometric) isolation criteria. Nevertheless, considering the third dimension, only 1/3 of the sample has no similar redshift neighbours \citep{2013A&A...560A...9A}. 

In this context the CIG represents an excellent sample to study the relation of galaxy properties on both local and large-scale environments. 

In the present work, we aim to identify and quantify the effects of the satellite distribution around a sample of CIG galaxies, as well as the effects of the Large Scale Structure. 
This study is organised as follows: in Sect.~\ref{Sec:data}, the sample and the data used are presented. 
The method to identify the potential satellite galaxies is described in Sect.~\ref{Sec:physical}. In Sect.~\ref{Sec:isolparam}, we describe the parameters used to quantify the environment. We present our results in Sect.~\ref{Sec:results} and the associated discussion in Sect.~\ref{Sec:discussion}. Finally a summary and the main conclusions of the study are presented in Sect.~\ref{Sec:con}. Throughout the study, a cosmology with $\Omega_{\Lambda 0} = 0.7$, $\Omega_{\rm{m} 0} = 0.3$, and $H_{0}=70$\,km\,s$^{-1}$\,Mpc$^{-1}$ is assumed.

%__________________________________________________________________

\section{The sample and the data} \label{Sec:data}

The CIG \citep{1973AISAO...8....3K} has been assembled with the requirement that no similar size galaxy $i$ with angular diameter $D_{i}$ between $1/4$ and $4$ times the apparent diameter $D_{P}$ of the primary CIG galaxy lies within $20\,D_{i}$ (Eqs.~\ref{Eq:kara2} and \ref{Eq:kara1}):

\begin{equation} \label{Eq:kara2}
\frac{1}{4} \,D_{P} \leq D_{i} \leq 4 \,D_{P}\quad;
\end{equation} 

\begin{equation} \label{Eq:kara1}
R_{iP} \geq 20 \,D_{i}\quad.
\end{equation} 

Until recently, most of the identifications and evaluations of large samples of isolated galaxies have been carried out using photometric data. The advent of the Sloan Digital Sky Survey \citep[SDSS;][]{2000AJ....120.1579Y,2011AJ....142...72E} has opened up the possibility to develop a detailed spectroscopic study of the environment of galaxies in the CIG.

Our starting sample is based on the CIG galaxies found in the ninth data release \citep[DR9;][]{2012ApJS..203...21A} of the SDSS. We focus our study on CIG galaxies with recession velocities $\varv \geq 1500$\,km\,s$^{-1}$ \citep{2007A&A...470..505V} so as to avoid an overwhelmed search for potential neighbours (the angular size on the sky for 1\,Mpc at a distance of 1500\,km\,s$^{-1}$ is approximately 2\fdg9). 
We then add the requirement that more than 80\% of the neighbours within a projected radius of 1\,Mpc possess a spectroscopic redshift in either the main galaxy sample \citep{2002AJ....124.1810S}, with magnitudes between $14.5~<~m_{r,\rm{Petrosian}}~<~17.77$, or in the Baryon Oscillation Spectroscopic Survey \citep[BOSS;][]{2013AJ....145...10D} which uses a new spectrograph \citep{2012arXiv1208.2233S} to obtain spectra of galaxies with $0.15 < z < 0.8$ and quasars with $2.15 < z < 3.5$, thus useful to reject background objects in our study. 
% The SDSS data are processed using automatic pipelines \citep{2011AJ....142...31B}.
To correct for redshift incompleteness in the field, we use the photometric redshift $z_p$ provided by the SDSS \citep[\normalfont{z} of the table \normalfont{Photoz}, for galaxies at magnitudes $m_{r} < 17.77$ according to][]{2013MNRAS.430..638S}. After a first rejection of neighbours with $z_p > 0.1$ as background galaxies, we select as potential companions neighbour galaxies with $|z_{\rm{CIG}}-z_p|~<~2.5\,z_{p,\rm{Err}}$ \citep{2011MNRAS.417..370G}, where $z_{\rm{CIG}}$ is the spectroscopic redshift of the CIG galaxy and $z_{p,\rm{Err}}$ is the photometric redshift error.

In order to evaluate the effects of the large scale environment, we follow a methodology similar to \citet{2013A&A...560A...9A}, searching for neighbours around 386 CIG fields completely covered by the SDSS within a physical projected radius of 3\,Mpc.

Model magnitudes in the $r$-band (the deepest images) are used in our study. Sizes are estimated from $r_{90}$, the Petrosian radius containing 90\,\% of the total flux of the galaxy in the $r$-band\footnote{\texttt{http://www.sdss3.org/dr9/algorithms/magnitudes.php}}, as explained in \citet{2013A&A...560A...9A}. Absolute magnitudes and stellar masses (see upper panel in Fig.~\ref{Fig:moster}) for both CIG galaxies and neighbours, are calculated by fitting the spectral energy distribution using the routine kcorrect \citep{2007AJ....133..734B}.

%______________________________________________________________

\section{Identification of physical companions} \label{Sec:physical}

To recover the physical satellites around the CIG galaxies, we first focus on the satellites which are within the escape speed of each CIG galaxy (Sect.~\ref{Sec:escape}). We also propose a more conservative method using the stacked Gaussian distribution of the velocity difference of the neighbours, with respect to the corresponding CIG galaxy, which gives an upper limit for the influence of the local environment (Sect.~\ref{Sec:gaussian}).

\subsection{Escape speed} \label{Sec:escape}

To identify the physical satellites which may have had a secular influence on the central CIG galaxy, we use the escape speed in order to select the physically bound satellite galaxies. The escape speed at a given distance reads:

\begin{equation} \label{Eq:velesc}
\varv_{\rm{esc}} = \sqrt{\frac{2GM_{P}}{R_{iP}}} \quad ,
\end{equation}
    
\noindent where $G$ is the universal gravitational constant, $M_{P}$ is the dynamical mass of the central CIG galaxy, and $R_{iP}$ is the distance between the neighbour $i$ and the primary galaxy $P$.

\begin{figure}
\begin{center}
\includegraphics[width=\columnwidth]{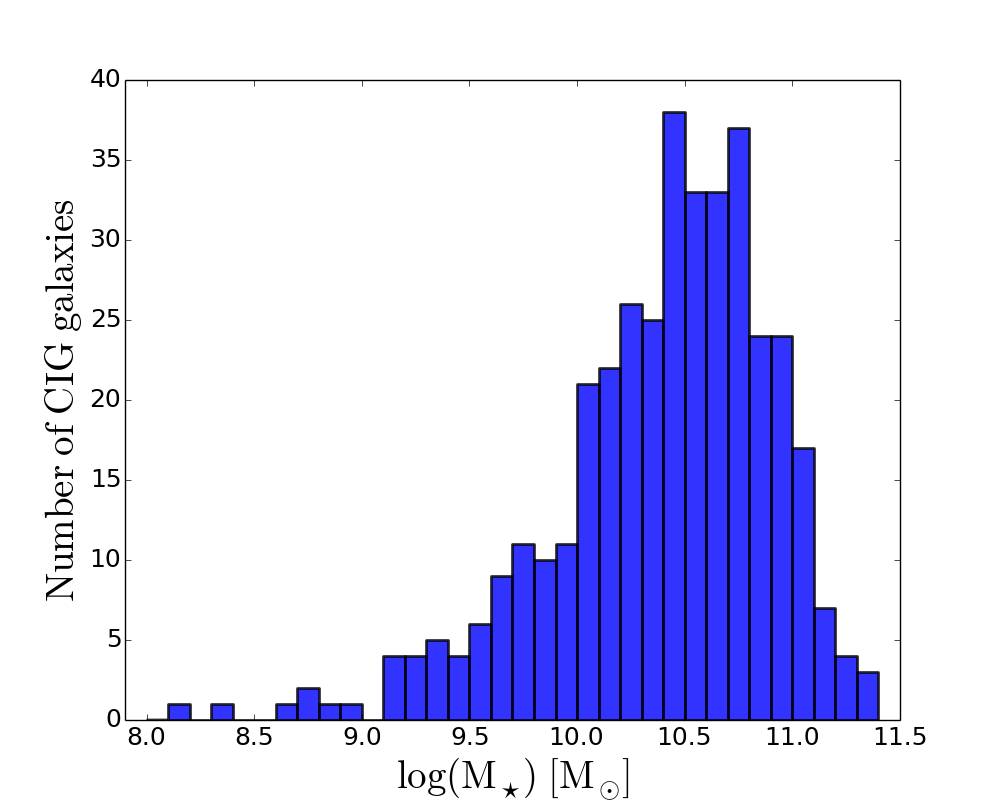} \\
\includegraphics[width=\columnwidth]{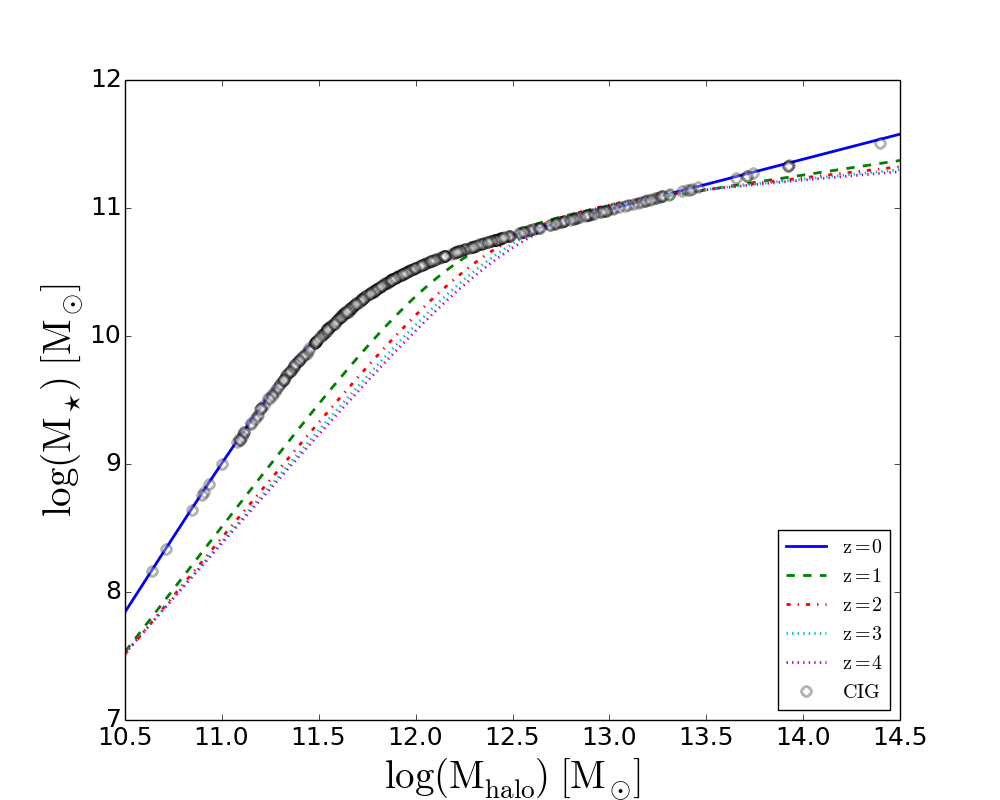} \\
\includegraphics[width=\columnwidth]{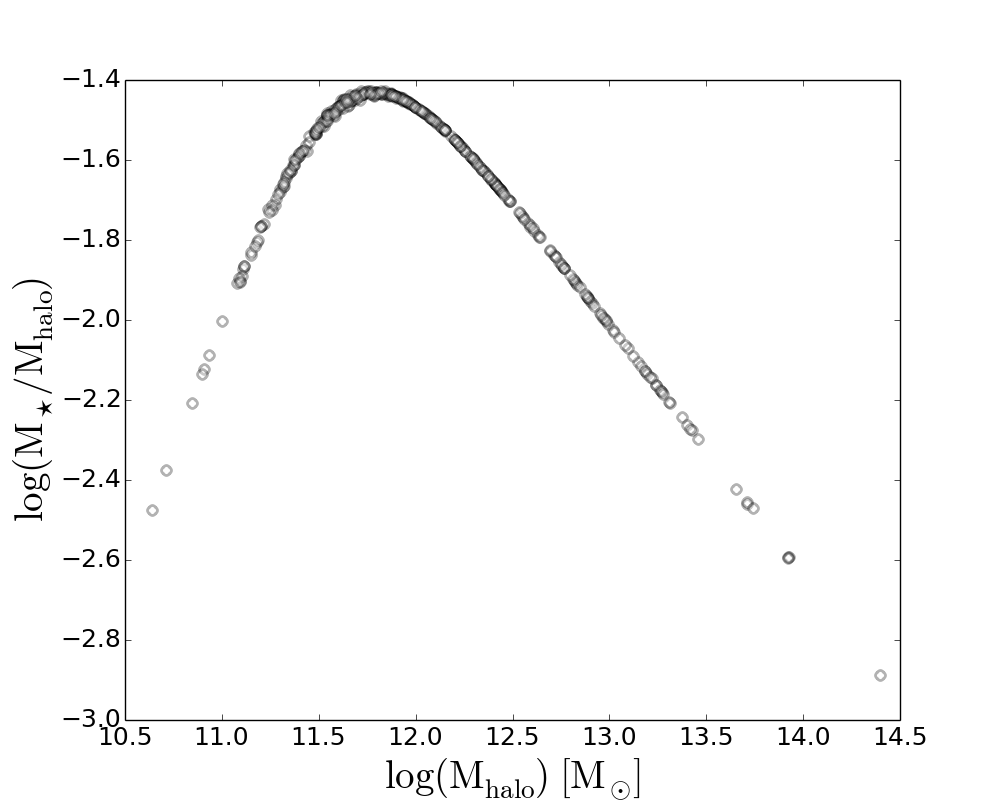}
\end{center}
\caption[SHM relation for central galaxies as a function of redshift]{{\it (upper panel):} Distribution of the stellar masses of the CIG galaxies. {\it (middle panel):} SHM relation for central galaxies as a function of redshift \citep[Figure 5 in][]{2013MNRAS.428.3121M}. Blue solid line, green dashed line, red point-dashed line, cyan pointed line, and magenta pointed line correspond to the SHM relation at redshift 0, 1, 2, 3, and 4, respectively. Black circles correspond to the parametrisation of the SHM relation for CIG galaxies.
{\it (lower panel):}  Logarithm of the SHM ratio for CIG galaxies.} \label{Fig:moster}
\end{figure}

In the upper panel of Fig.~\ref{Fig:moster}, the distribution of the stellar masses of the CIG galaxies is presented. The logarithm of the stellar mass spans 8.1--11.4\,$\rm M_\odot$, with a peak towards 10.5\,$\rm M_\odot$. The dynamical masses of the CIG galaxies are estimated from their stellar masses following the parametrisation in \citet{2013MNRAS.428.3121M}, including redshift evolution (see their Eq.~2). The result of the Stellar-to-Halo Mass (SHM) parametrisation for CIG galaxies is shown in middle and lower panels in Fig.~\ref{Fig:moster}.

The escape separation should be computed using the 3-dimensional space between the central CIG galaxy and its neighbours. Unfortunately we do not have this information and the use of projected separations, as well as 1-dimensional line-of-sight velocities, would lead to an overestimation of the number of physical companions. To correct for this bias we first assume an isotropic velocity distribution (see also next section). Under this hypothesis, the 1-dimensional, line-of-sight velocity is related to the 3-dimensional velocity by a scale factor $\sqrt{3}$. Similarly, the projected separation should be multiplied by a factor $\frac{\sqrt{3}}{\sqrt{2}}$ to approximate 3D separation. Although these approximations may not be totally exact for a given CIG galaxy, they nevertheless represent a useful first step towards a full 3D characterisation of the environment of galaxies.

\begin{figure}
\begin{center}
\includegraphics[width=\columnwidth]{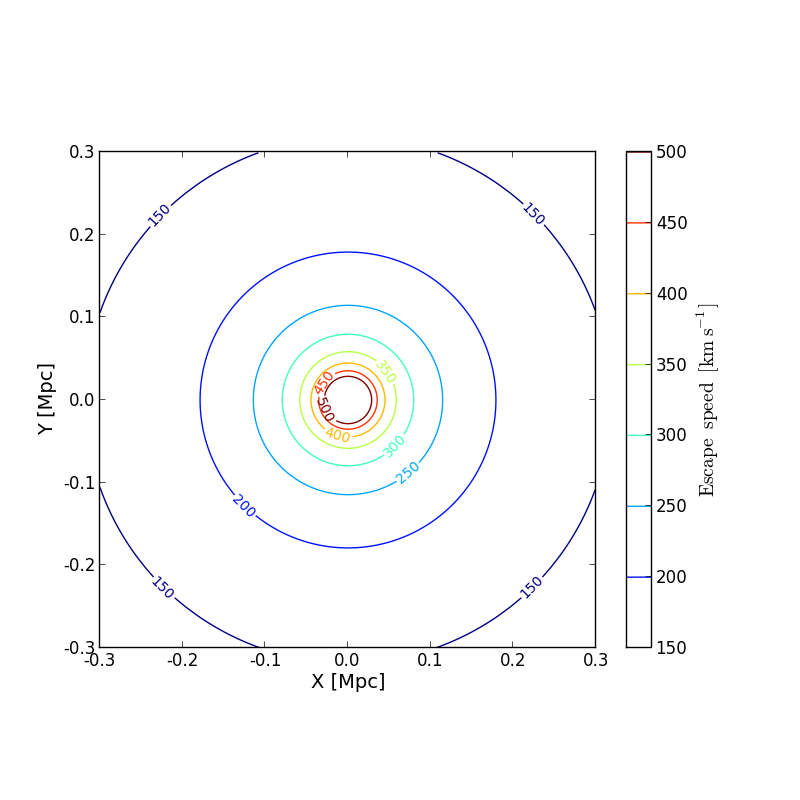} \\
\includegraphics[width=\columnwidth]{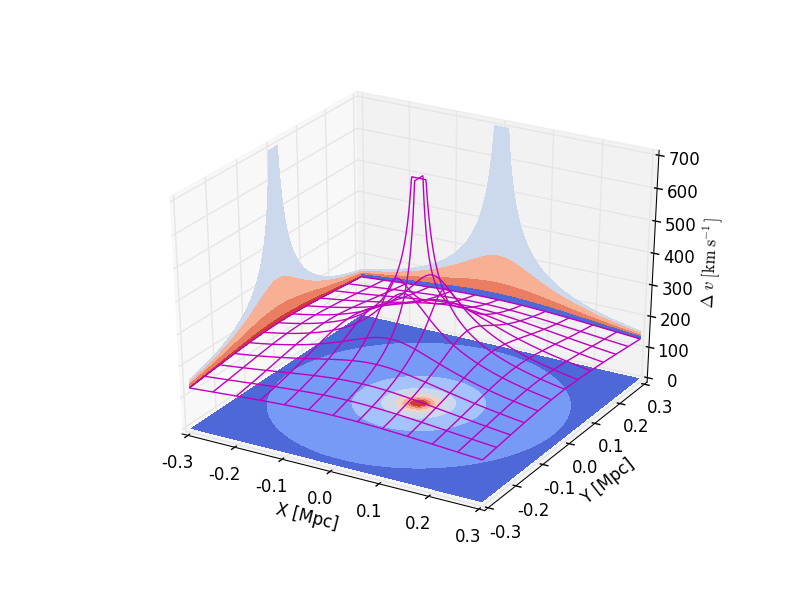}
\end{center}
\caption[Escape speed method]{
{\it (upper panel):} 2-Dimensional escape speed schema in celestial coordinates space.
The physically linked associations correspond to neighbour galaxies at distance to the central CIG galaxy and $|\Delta\,\varv|$ less than the corresponding value according to circle lines. The levels are calculated for a typical stellar mass of $10^{10.5}$\,$\rm M_\odot$, translating into a dynamical mass of $10^{11.9}$\,$\rm M_\odot$.
{\it (lower panel):} 3-Dimensional escape velocity schema in a redshift-space diagram (line-of-sight velocity versus projected distance). The physically linked associations correspond to neighbour galaxies under the 'trumpet' surface. The levels are calculated for a typical stellar mass of $10^{10.5}$\,$\rm M_\odot$, translating into a dynamical mass of $10^{11.9}$\,$\rm M_\odot$.} \label{Fig:caustic}
\end{figure}

Several levels of the escape speed are shown in the upper panel of Fig.~\ref{Fig:caustic}, as a function of the projected distance, up to 0.3\,Mpc. The levels are calculated for a typical stellar mass of $10^{10.5}$\,$\rm M_\odot$, translating into a dynamical mass of $10^{11.9}$\,$\rm M_\odot$. In the lower panel, we show the characteristic trumpet shape \citep[caustic][]{1987MNRAS.227....1K, 1995PhR...261..271S,1997ApJ...481..633D} under which the satellite galaxies would be captured. Above the caustic, the neighbour galaxies possess a velocity sufficient to evade the gravitational attraction of the primary galaxy and are not captured, although fly-by interactions may influence the structure and evolution of the primary galaxy.

Using the escape speed, the satellite galaxies considered physically bound with their corresponding CIG galaxy, are all neighbours satisfying the condition $|\Delta\,\varv| \leq \sqrt{\frac{2GM_{P}}{3\frac{\sqrt{3}}{\sqrt{2}}R_{iP}}}$\,km\,s$^{-1}$ and lying at a distance lower than 0.3\,Mpc.

\subsection{Gaussian distribution of physical satellites} \label{Sec:gaussian}

Some galaxies may pass nearby a primary CIG galaxy, but with a velocity so high that they interact once with the CIG galaxy and then leave. 
To take into account the potential effect of fly-by encounters, we develop a more conservative method to recover most of the galaxies which have interacted with the CIG galaxies. We do so by stacking all the primaries and their satellites in order to obtain statistically robust results.

\begin{figure}
\begin{center}
\includegraphics[width=\columnwidth]{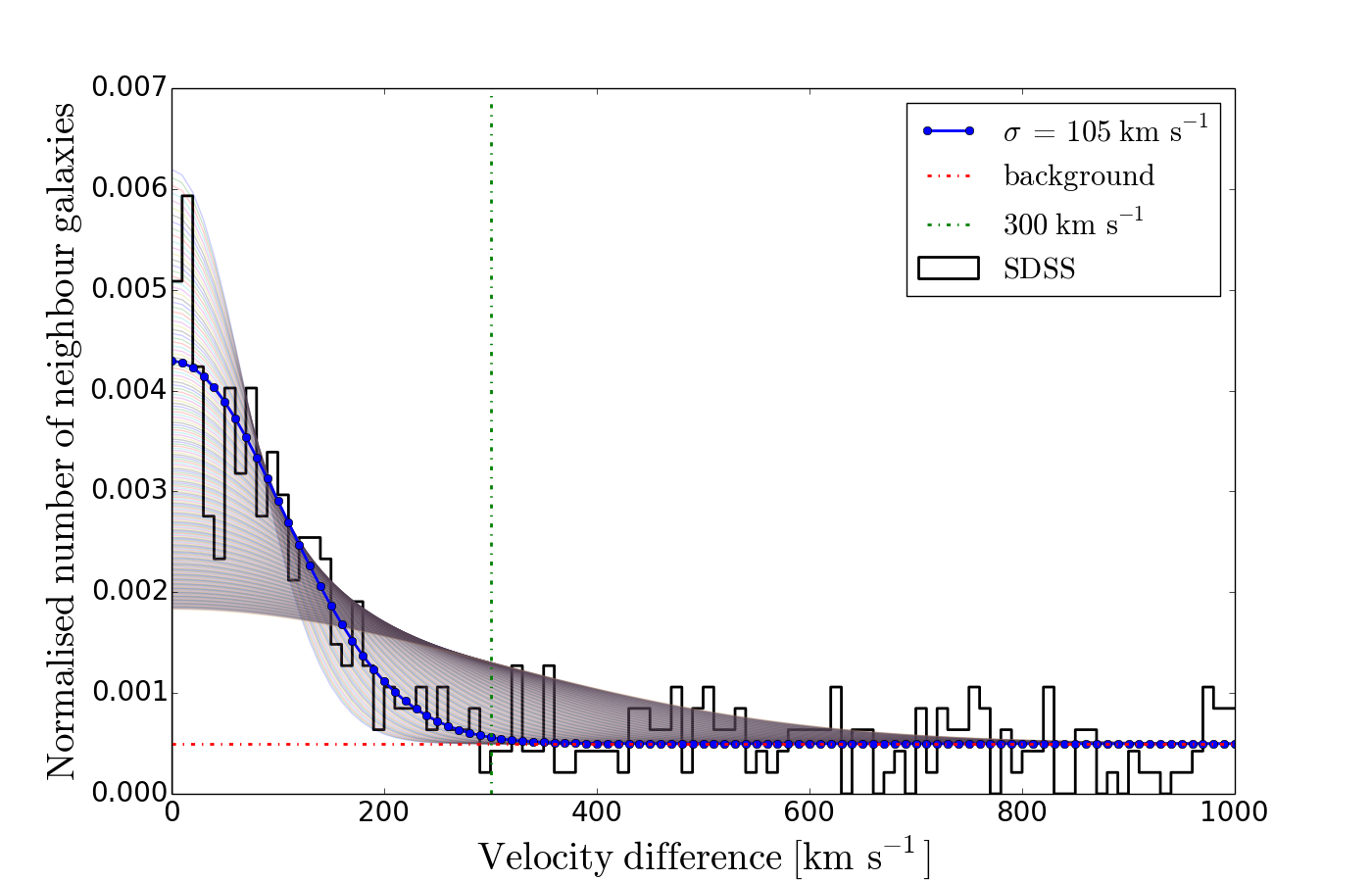} \\
\includegraphics[width=\columnwidth]{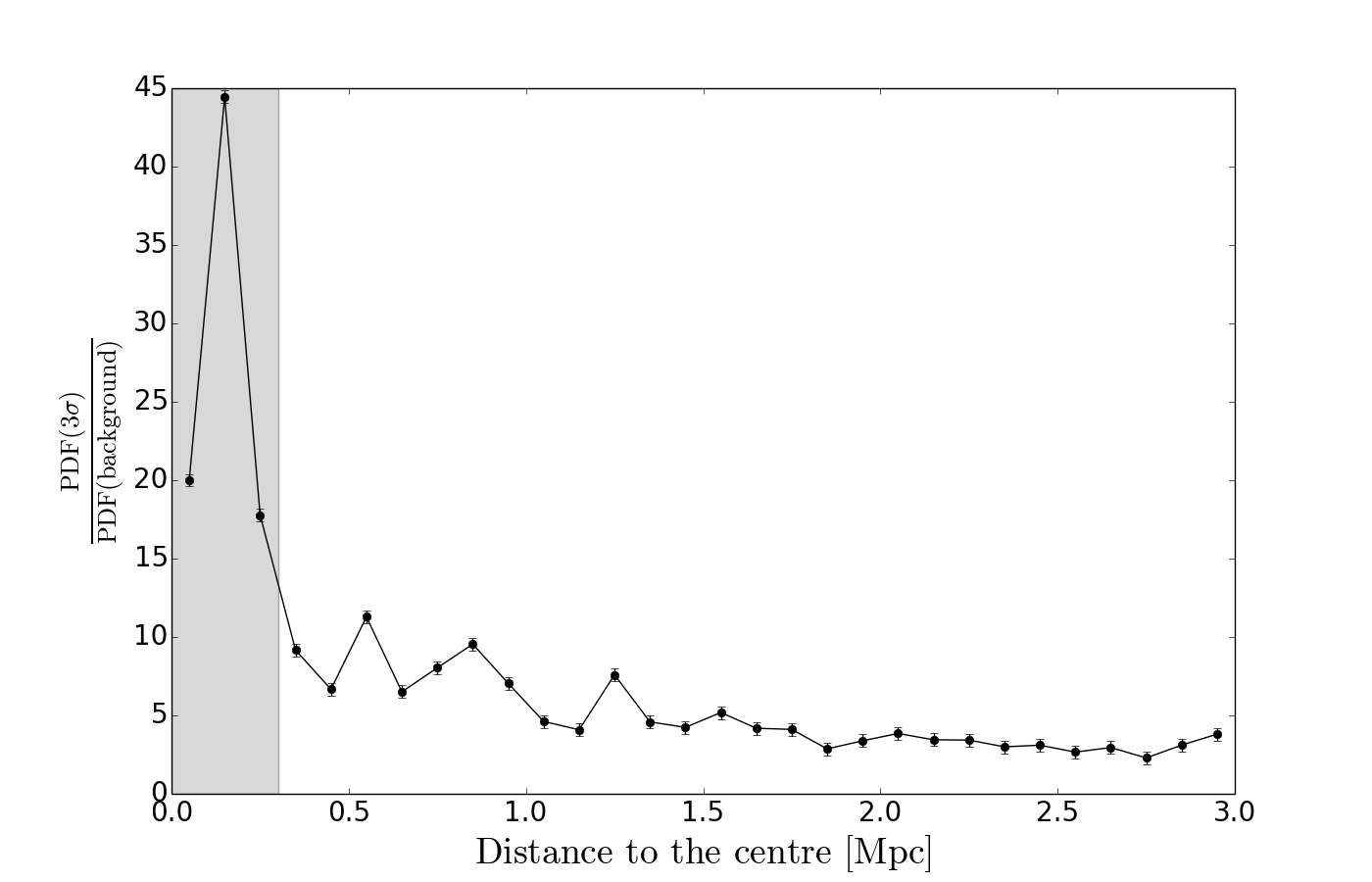}
\end{center}
\caption[Gaussian distribution of physical satellites]{{\it (upper panel):} Absolute values of the line-of-sight velocity difference between neighbours and the central CIG galaxy: $|\Delta\,\varv|$ distribution obtained by stacking 411 CIG fields within 1\,Mpc field radius (black histogram), and corresponding Gaussian distribution fits for $\sigma$ between 70 and 300\,km\,s$^{-1}$ (grey curves) with the best fit (blue curve). The Gaussian fit has been done within $|\Delta\,\varv| = 300$\,km\,s$^{-1}$ (vertical line) and considering as a zero point the flat continuum distribution of background neighbours (horizontal line).
{\it (lower panel):} Probability Density Function (PDF) for neighbour galaxies peaking at $|\Delta\,\varv|~=~0$\,km\,s$^{-1}$ over PDF for the background flat population selected in the interval $1000 < |\Delta\,\varv| < 3000$\,km\,s$^{-1}$, as a function of the distance to the central CIG galaxy. The inner 0.3\,Mpc are shaded.} \label{Hist:diffvelold}
\end{figure}

In the upper panel of Fig.~\ref{Hist:diffvelold}, we show the distribution of the absolute values of the radial velocity difference between the projected neighbour galaxies and the central CIG galaxies ($\Delta\,\varv = \varv_{\rm neigh} - \varv_{\rm CIG}$). Two components appear clearly in the figure. The first component is a flat continuum distribution of foreground/background neighbours, extending to Mpc scales, and related to the LSS distribution of galaxies. The second component is the over-abundance of neighbour galaxies peaking at $|\Delta\,\varv|~=~0$\,km\,s$^{-1}$; most of those would be dynamically related to the central CIG galaxies.  
In order to estimate the standard deviation, $\sigma$, of the distribution, we first estimate the median level of the background between 300 and 1000\,km\,s$^{-1}$, and remove it. A Gaussian distribution appears for velocity differences minor than 300\,km\,s$^{-1}$. We vary $\sigma$ between 70 and 300\,km\,s$^{-1}$ and use a $\chi^2$ fitting minimisation to obtain the standard deviation of the satellite distribution: $\sigma = 105$\,km\,s$^{-1}$. Consequently, the $3 \sigma$ limit is at 315\,km\,s$^{-1}$. The neighbour galaxies with $|\Delta\,\varv| \leq$~315\,km\,s$^{-1}$ show a substantial liability to gather in the inner 0.3\,Mpc around the CIG galaxies (see the lower panel of Fig.~\ref{Hist:diffvelold}). This dynamical link is also confirmed by the constancy of the standard deviation for radii lower than 0.3\,Mpc (see the upper panel of Fig.~\ref{Fig:excess}, and the associated analysis in Sect.~\ref{subsec:LocalvsLSS}).
To be very conservative and recover 99.7\% of the physically linked companions, we consider that all neighbours within $|\Delta\,\varv| \leq 3\,\sigma$ may be physically bound with their corresponding CIG galaxy.

Hence, the satellite galaxies selected by the Gaussian distribution are all neighbour galaxies with $|\Delta\,\varv| \leq$~315\,km\,s$^{-1}$ and lying at a distance lower than 0.3\,Mpc. This method provides an upper limit on the quantification of the local environment, since more galaxies will be considered as satellites with respect to those selected following the escape speed method.

%______________________________________________________________

\section{Quantification of the environment} \label{Sec:isolparam}

In order to quantify the isolation degree of the CIG galaxies, we use two complementary parameters: the tidal strength $Q$ that the neighbours produce on the central galaxy \citep{2007A&A...472..121V,2013MNRAS.430..638S,2013A&A...560A...9A}, and the projected density $\eta_{k}$ (Eqs.~\ref{Eq:Q2012tot} and \ref{Eq:etak}) of neighbour galaxies considered in this study.
 
\subsection{Tidal strength parameter}

The tidal strength parameter is defined as:
\begin{equation} \label{Eq:Qip}
Q_{iP} \equiv \frac{F_{\rm{tidal}}}{F_{\rm{bind}}} 
\propto {\frac{M_{i}}{M_{P}}} \left(\frac{D_{P}}{R_{iP}}\right)^3\quad,
\end{equation} 
where $M_i$ and $M_P$ are the stellar masses of the neighbour and the principal galaxy, respectively, $D_{P}$ the apparent diameter of the principal galaxy, and $R_{iP}$ the projected physical distance between the neighbour and principal galaxy. We use the apparent diameter $D_P=2\,r_{90}$ scaled by a factor 1.43 \citep{2013A&A...560A...9A} to match the definition of diameter used in the literature \citep[projected major axis of a galaxy at the 25\,mag\,arcsec$^{-2}$ isophotal level or $D_{25}$,][]{2007A&A...472..121V}.
The total tidal strength is then defined as:
\begin{equation} \label{Eq:Q2012tot}
Q = \log \left(\sum_i Q_{iP}\right) \quad .
\end{equation}
The logarithm of the sum of the tidal strength created by all the neighbours in the field is a dimensionless estimation of the gravitational interaction strength \citep{2007A&A...472..121V}. The greater the value of $Q$, the less isolated from external influence the galaxy, and vice-versa.

\subsection{Projected number density parameter}

To characterise the LSS around the CIG galaxies, we also define the projected number density parameter 
\citep{2007A&A...472..121V,2013A&A...560A...9A}, as follows:
\begin{equation} \label{Eq:etak}
\eta_{k, \rm LSS} \propto \log \left(\frac{k - 1}{V(r_k)}\right)\quad,
\end{equation} 
where $V(r_k) = \frac{4}{3}\,\pi\,r_k^3$ and $r_k$ is the projected physical distance to the $k^{\rm th}$ nearest neighbour, with $k$ equal to 5, or less if there were not enough neighbours in the field. The farther the $k^{\rm th}$ nearest neighbour, the smaller the projected number density $\eta_{k, \rm LSS}$.

%______________________________________________________________

\section{Results} \label{Sec:results}
\subsection{Spectroscopic identification of physical satellites around galaxies in the CIG} \label{Sec:results-sat}

Out of the 386 CIG galaxies considered in this study, 340 have no physically linked satellites, which represents 88\% of the sample. Among the 46 CIG galaxies with at least one physical companion within its escape speed boundary, 36 have one satellite, nine have two satellites, and one has three satellites (CIG 771). There is no CIG galaxy with more than three physically linked satellites. The values of the tidal forces exerted by these satellites on the CIG galaxies are listed in columns 2, 3, 4, and 5 of Table~\ref{Tab:isolparam}.

Following the more conservative Gaussian distribution of physical satellites around the CIG galaxies leads to upper limits. Out of the 386 CIG galaxies, 327 (85\% of the sample) have no physical companion within a projected distance of 0.3\,Mpc. Out of the remaining 59 CIG galaxies (15\%), 46, 11, and 2 CIG galaxies (CIG\,237 and CIG\,771) are in interaction with one, two, and three physical companions, respectively.

\begin{figure*}
\begin{center}
\includegraphics[width=\textwidth]{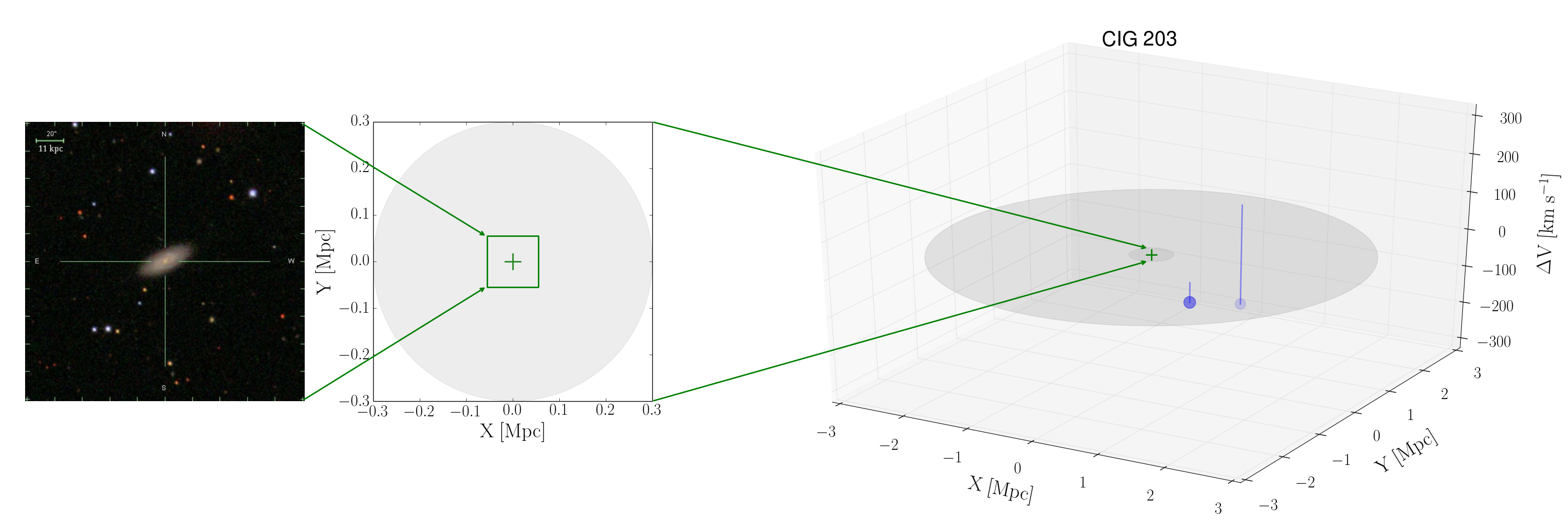}
\includegraphics[width=\textwidth]{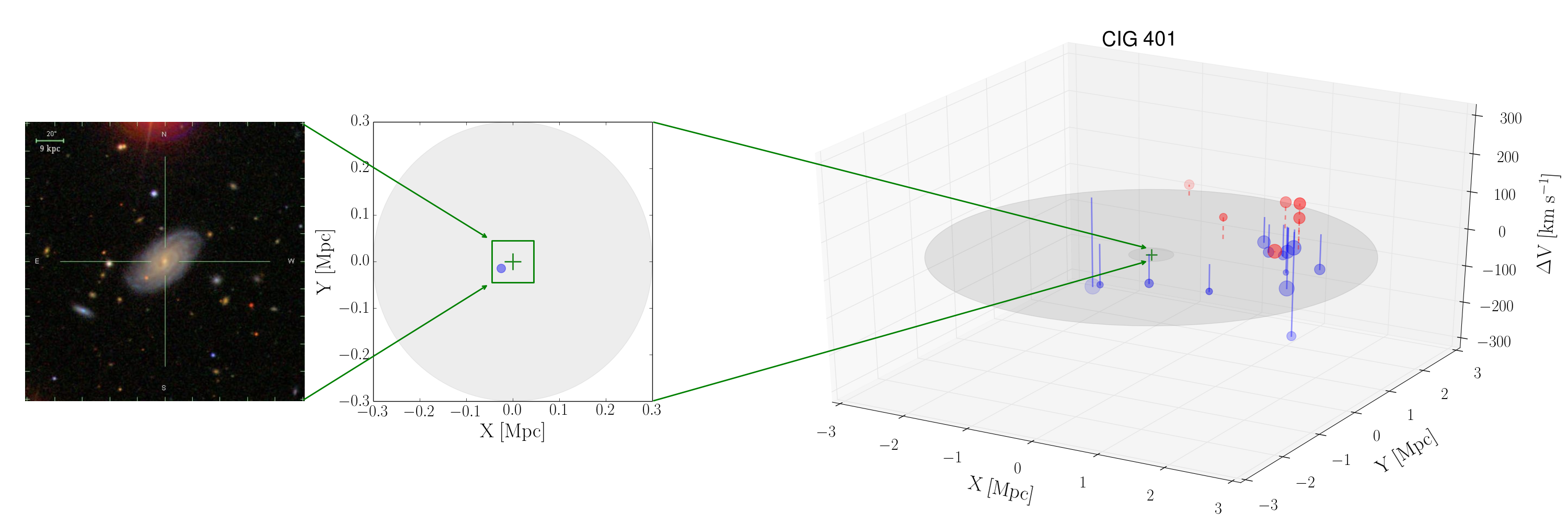}
\includegraphics[width=\textwidth]{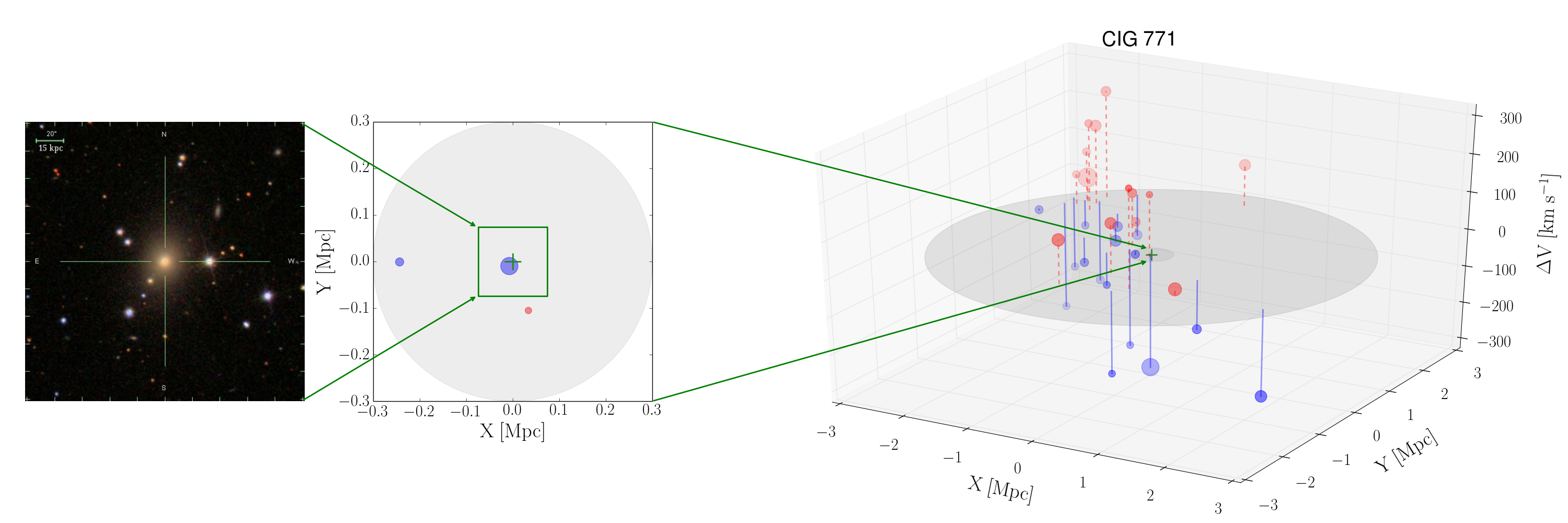}
\end{center}
\caption[3-Dimensional view of the environment]{SDSS three-colour images (left, with a field of view of 3\arcmin3, North is up, East is to the left), representation in 2D (center, with a field of view of 0.6\,Mpc, North is up, East is to the left), and representation in 3D (right, with a field of view of 6\,Mpc) of the environment of the galaxies CIG\,203 (upper panel), CIG\,401 (middle panel), and CIG\,771 (lower panel).
The central green pluses correspond to the locations of the primary CIG galaxies. Grey disks at 0.3 and 3\,Mpc boundaries represent the areas to study the local environment and the LSS, respectively. The sizes of the symbols are proportional to the diameters of the neighbours (not at scale), in red or blue according to blueshift or redshift of the neighbours. Vertical lines indicate the projection of the neighbours in the 2-Dimensional plane. Solid green lines indicate the scale relations between the three different representations.} \label{Fig:charts}
\end{figure*}

Examples of the environment for three CIG galaxies are shown in Fig.~\ref{Fig:charts}. CIG\,203 has no physically bound companions. On the other hand, CIG\,401 and CIG\,771 are linked locally with companions that are caught under their gravitational influence.

\subsection{Correction for the redshift incompleteness}

As mentioned in Sect.~\ref{Sec:data}, the sample used in this study was selected for CIG fields with at least 80\% redshift completeness (the mean completeness of the sample is 92.5\%). Therefore, to correct for this incompleteness we consider photometric redshifts to compensate the estimation of the local environment. The upper limit on the local tidal strength is then calculated considering the potential companions at the same distance as their corresponding CIG galaxy, i.e., the least favourable case for the isolation degree.

This correction is applied to the already conservative Gaussian distribution selection of physical satellites, introducing no change for 356 CIG galaxies (92\% of the sample). Out of the remaining 30 CIG galaxies, 22 CIG galaxies without satellite acquire one, six CIG galaxies pass from one to two satellites, and one CIG galaxy passes from two to three satellites. Only one CIG galaxy, CIG\,626, gains more than one possible satellite, passing from none to three possible satellites.

The inclusion of one missing redshift galaxy as a potential companion increases the tidal strength by a mean value of 13\%, with respect to the tidal strength generated by the spectroscopic satellites. The most unfavourable cases occur for CIG\,278 and CIG\,495 where the effect of the missing galaxy amounts to 64\% and 83\%, respectively.

\subsection{Large Scale Structure} \label{Sec:isolparamresults}

To quantify the large scale environment around the CIG galaxies, we use the two isolation parameters defined in Sect.~\ref{Sec:isolparam}. The parameter $Q_{\rm LSS}$ was calculated taking into account all companions within 3\,Mpc and $|\Delta\,\varv| < 315$\,km\,s$^{-1}$ to provide the sum of the tidal strengths exerted on the CIG galaxies. The parameter $\eta_{k, \rm LSS}$ accounts for the number density of companions within $|\Delta\,\varv| < 315$\,km\,s$^{-1}$ and projected at the distance of the $5^{\rm{th}}$ nearest neighbour with respect to the CIG galaxy. Only 10 CIG galaxies (less than 3\% of the sample) are farther away than 3\,Mpc from any other galaxy with a SDSS measured spectrum.

\begin{figure}
\begin{center}
\includegraphics[width=\columnwidth]{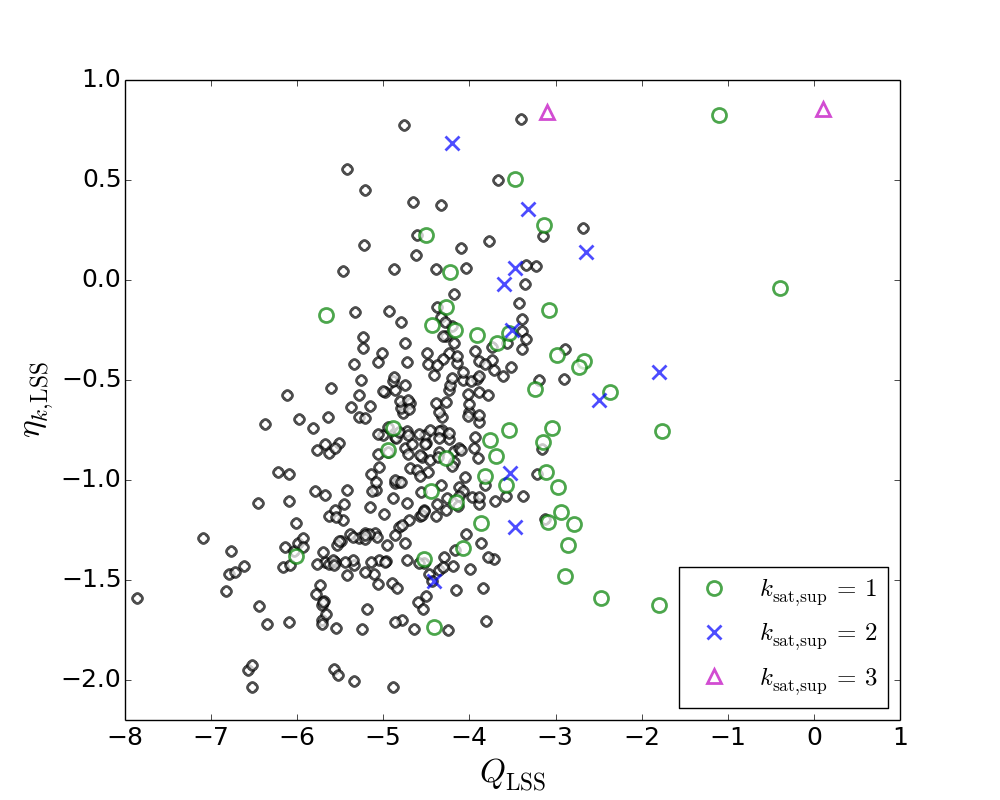} 
\end{center}
\caption[LSS isolation parameters]{Projected number density $\eta_{k, \rm LSS}$ versus tidal strength $Q_{\rm LSS}$ diagram for the LSS.
CIG galaxies with one, two, and three dynamically linked satellites are depicted by green circles, blue crosses, and magenta triangles, respectively.} \label{Fig:isolparam}
\end{figure}

In Fig.~\ref{Fig:isolparam}, the projected number density is shown versus the tidal strength parameter. Due to the logarithmic definition of the parameters, the figures appear to span several orders in magnitude (3\,dex for $\eta_{k, \rm LSS}$ and 7\,dex for $Q_{\rm LSS}$) showing the large scatter in the environments found around the CIG galaxies. The results of the quantifications of the environment are listed in columns 6, 7, and 8 of Table~\ref{Tab:isolparam}. CIG galaxies with physically bound satellites tend to show greater values of $Q_{\rm LSS}$ but not $\eta_{k, \rm LSS}$.

The large scale environment is graphically exemplified around three CIG galaxies in the right column of Fig.~\ref{Fig:charts}. There is a sparse population of galaxies in the LSS around CIG\,203, while CIG\,401 and CIG\,771 show a much more crowded LSS within 3\,Mpc.

\subsection{Local environment versus large scale environment} \label{subsec:LocalvsLSS}

To evaluate the role of the physically bound satellites with respect to the large scale environment, we compare the magnitudes of the sum of the tidal forces produced by the physical companions to the sum of the tidal forces engendered by all the galaxies in the LSS. When at least one physical companion is present near a CIG galaxy, its effect largely dominates (usually more than 90\%) over the tidal forces generated by the LSS. This effect is clearly visible in Fig.~\ref{Fig:isolparam}. The ratios $Q_{\rm sat} / Q_{\rm LSS}$ and $Q_{\rm sat, sup} / Q_{\rm LSS}$ are tabulated in columns 9 and 10 of Table~\ref{Tab:isolparam}, respectively.

\begin{figure}
\begin{center}
\includegraphics[width=\columnwidth]{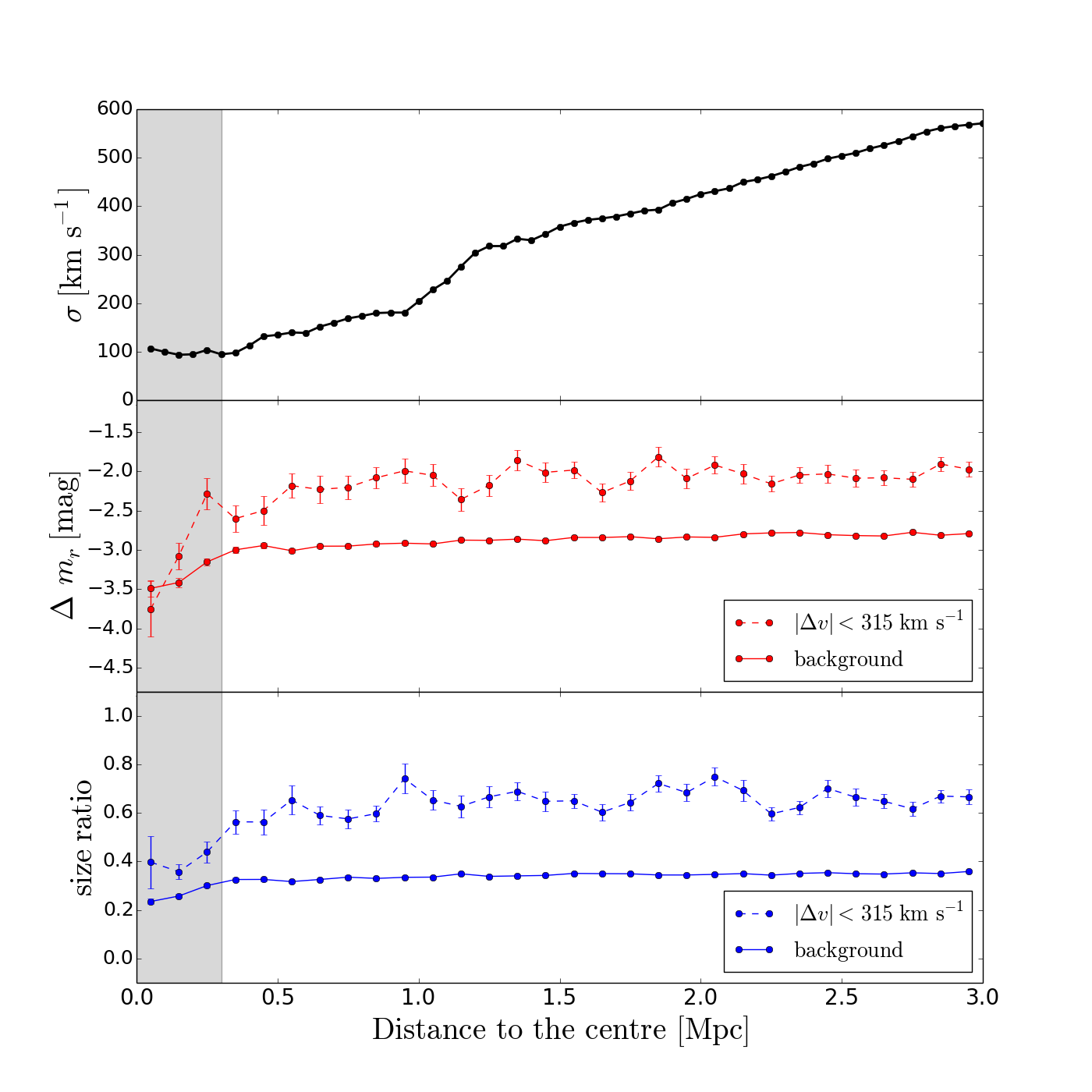}
\end{center}
\caption[Characterisation of satellites versus the background population]{{\it (upper panel):} Standard deviation $\sigma$ of the Gaussian distribution fitting for $|\Delta\,\varv|$ as a function of the distance to the central galaxy. 
{\it (middle panel):} Magnitude difference ($\Delta\,m_{r} = m_{r}^{P} - m_{r}^{i}$ between neighbour $i$ and its corresponding primary CIG galaxy $P$) for satellites ($|\Delta\,\varv| \leq 315$\,km\,s$^{-1}$, red dashed line) and for background galaxies ($315 < |\Delta\,\varv| < 3000$\,km\,s$^{-1}$, red solid line) as a function of the distance to the central CIG galaxy.
{\it (lower panel):} Size ratio ($\frac{D_{i}}{D_{P}}$ between neighbour $i$ and its corresponding primary CIG galaxy $P$) for satellites ($|\Delta\,\varv| \leq 315$\,km\,s$^{-1}$, blue dashed line) and for the background population ($315 < |\Delta\,\varv| < 3000$\,km\,s$^{-1}$, blue solid line) as a function of the distance to the central CIG galaxy. The inner 0.3\,Mpc are shaded in the three panels.} \label{Fig:excess}
\end{figure}

It seems that there is a natural distinction between the physically bound satellites and the LSS. This dichotomy appears for instance if we plot the standard deviation of a Gaussian fitting as a function of the projected distance (see the upper panel of Fig.~\ref{Fig:excess}). Up to $\sim0.3$\,Mpc, most of the galaxies are linked to their host. If the systems have been in interaction long enough, they are relaxed and the velocity differences are virialised. This appears as a plateau in the inner $\sim0.3$\,Mpc, with a constant standard deviation $\sigma \approx 105$\,km\,s$^{-1}$. At larger distances, the standard deviation monotonically increases due to the rising fraction of the LSS galaxies enclosed.

The physically captured satellite galaxies are typically 1.5\,dex fainter than the magnitude of galaxies lying farther away (see the middle panel of Fig.~\ref{Fig:excess}). In average, the galaxies in the inner $\sim 0.3$\,Mpc are also smaller (about $0.4 \times D_P$) compared to galaxies which are not satellites ($\sim 0.7 \times D_P$).

The connection between the CIG and the LSS is revealed by comparing the apparent magnitudes and sizes of the galaxies with $|\Delta\,\varv| < 315$\,km\,s$^{-1}$ to the ones of the galaxies outside this limit (see middle and lower panels of Fig.~\ref{Fig:excess}, respectively). Galaxies with similar velocities have magnitudes and sizes closer to the ones of CIG galaxies, compared to background and foreground galaxies (defined by galaxies with recession velocities in the range $315 < |\Delta\,\varv| < 3000$\,km\,s$^{-1})$, disclosing the association between the CIG galaxies and their surrounding LSS.

%______________________________________________________________

\section{Discussion} \label{Sec:discussion}

\subsection{The construction of the CIG}

In a previous work \citep{2013A&A...560A...9A} we revised the CIG isolation criterion using both photometry and spectroscopy from the SDSS-DR9. We found that the 16\% of the CIG galaxies considered in the spectroscopic study do not pass the CIG isolation criterion. There may be a population of very close physical satellites which may have a considerable influence on the evolution of the central CIG galaxy. Therefore, one of the aims of the present study is to characterise such a population of satellites.

As shown in Sect.~\ref{Sec:results-sat}, about 12\% (and up to 15\%) of the CIG galaxies have physically bound satellites. To seek why these systems are included in the CIG, it is worth to recall that the CIG has been constructed visually, on photographic material \citep{1973AISAO...8....3K}. Unfortunately, the sample of neighbour galaxies inspected originally is not available. Nevertheless, a revision has been carried out by \citet{2007A&A...470..505V} on the same original material (Palomar Observatory Sky Survey, POSS), providing a catalogue of approximately 54,000 neighbours. By comparing the physically bound satellites found in the SDSS to this POSS-based catalogue, we should be able to point out some drawbacks due to the use of photographic plates and the nearly total lack of redshift availability, forty years ago.

\begin{figure}
\begin{center}
\includegraphics[width=\columnwidth]{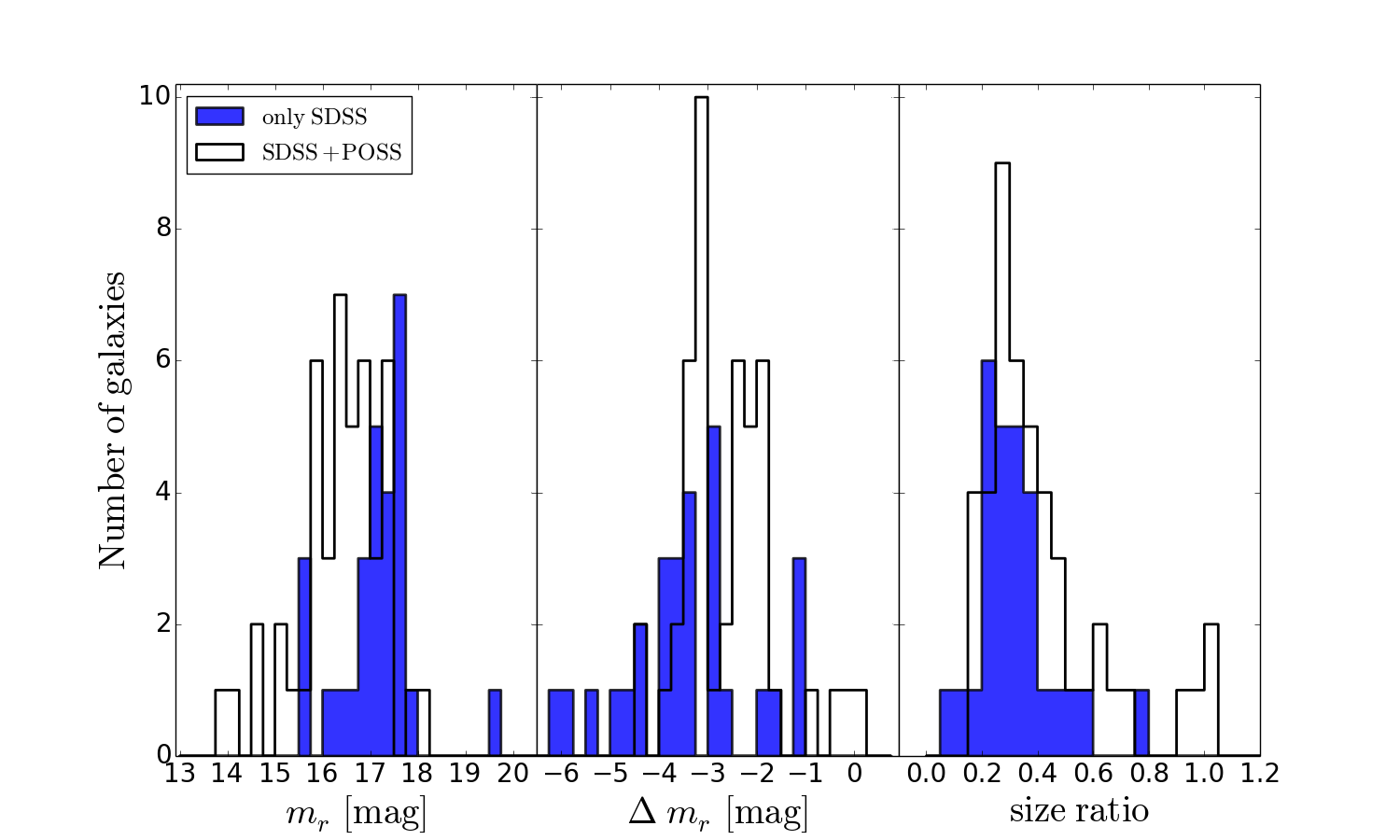}
\end{center}
\caption[Comparison of satellites with \citet{2007A&A...470..505V}]{Distribution of apparent magnitude (left panel), apparent magnitude difference (central panel), and size ratio (right panel) for physical linked satellites. Distributions for satellites which were not identified in the POSS \citep{2007A&A...470..505V} are represented by blue histograms.} \label{Fig:POSS}
\end{figure}

In the left panel of Fig.~\ref{Fig:POSS}, we show that the SDSS identification of satellites goes, in general, deeper than the POSS. Indeed, \citet{2007A&A...470..505V} recover neighbour galaxies brighter than $B = 17.5$. The slight overlap of magnitudes between the two distributions is due to the non-linearity of the photographic material, as well as the varying zero-point from field to field in the POSS calibration \citep{2007A&A...470..505V}. Equally, in the central panel of Fig.~\ref{Fig:POSS}, we see that the POSS search for companions misses the faintest galaxies, with respect to the magnitudes of the primary CIG galaxies.

Nonetheless, \citet{1973AISAO...8....3K} did not use any magnitude criterion to search for companions, and used only the apparent diameters of galaxies instead. In the right panel of Fig.~\ref{Fig:POSS}, it is shown that about half of the physical companions missed by the POSS have diameters smaller than one fourth of the diameter of their corresponding CIG galaxy, and were therefore not considered by the CIG isolation criteria. In fact, 23\% of the missing physical companions are dwarf galaxies discarded by the original study. For instance, in the case of CIG\,771, none of its three physically linked galaxies would violate the CIG isolation criterion. Nevertheless, its closest satellite, at 12 kpc and with a velocity difference of 230\,km\,s$^{-1}$, tackled by the present study may have a considerable influence on the evolution of CIG\,771. Redshift surveys are mandatory in order to distinguish small, faint, physically bound satellites from a background projected galaxy population.

\subsection{Identification of satellites}

Due to the spectroscopic redshift limit of the SDSS, analysing the satellite population around isolated galaxies is a challenge. We have some limitations to take into account. We are only able to detect bright neighbours (M\,31 or M\,33 like galaxies) and the brightest dwarfs (Large Magellanic Cloud and Small Magellanic Cloud like satellites) around most central galaxies. Nevertheless, the spectroscopic catalogue of the SDSS is complete to $m_r~<~17.77$\,mag, so we are always able to detect neighbours within $\Delta~m_r \leq 2$\,mag, even for the faintest CIG galaxy.

The spatial location of satellites with respect to their primary galaxies is uncertain due to redshift space distortions and projection effects. We follow a very conservative approach and use the projected separation between the neighbour and the CIG galaxy, providing a lower limit on the 3D distance between the two galaxies, which increases the number of potentially linked satellites taken into account by the escape speed and Gaussian distribution selections. This will translate on conservative higher limits for the isolation parameters.

In addition, the redshifts only account for the radial (line of sight) component of the peculiar velocities of the galaxies. Consequently, the velocity difference $\Delta\,\varv$ supplies also a lower limit and exaggerates the number of physically related companions, in particular for the escape speed selection method. Nevertheless, the escape velocity method selects satellites minimising the effect of background objects, although there is an uncertainty about the total dynamical mass of the primary galaxy. 

On the other hand, the Gaussian distribution method selects as satellites all neighbours within $|\Delta\,\varv| \leq 3 \sigma$ and at projected physical distances to the central galaxy $d \leq 0.3$\,Mpc. The $3 \sigma$ cut ensures that we recover more than 99.7\% of the physically associated satellites. This method includes also a fraction of fly-by encounters that may have an influence on the evolution of the primary galaxies. However, this is a compromise because, at the same time, it incorporates galaxies which are more likely located at 4.5\,Mpc from the primary galaxy rather than at the same distance and with a velocity difference $\Delta\,\varv = 315$\,km\,s$^{-1}$ (following the lineal approximation of the Hubble law $v = H_0 D$). This explains why the Gaussian distribution provides an upper limit to the escape speed selection.

\subsection{Local and large scale environments around CIG galaxies}

\begin{figure*}
\begin{center}
\includegraphics[width=0.19\textwidth]{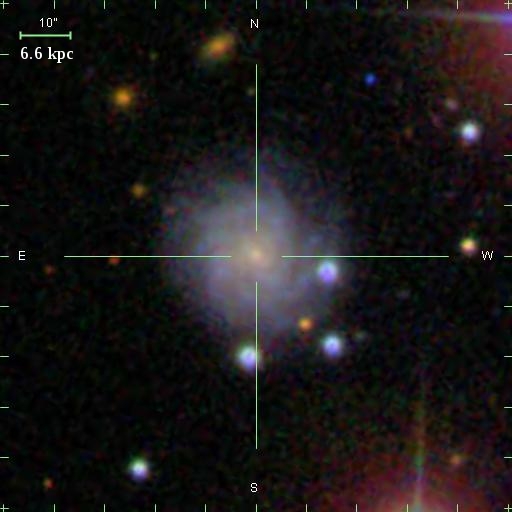}
\includegraphics[width=0.19\textwidth]{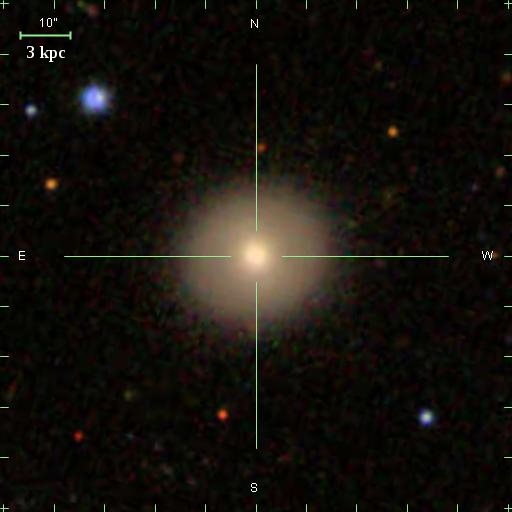}
\includegraphics[width=0.19\textwidth]{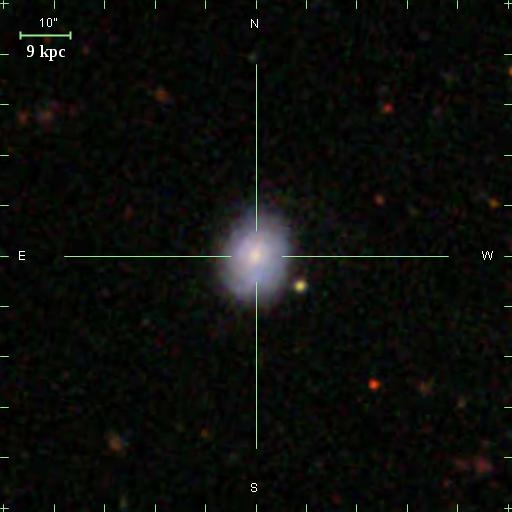}
\includegraphics[width=0.19\textwidth]{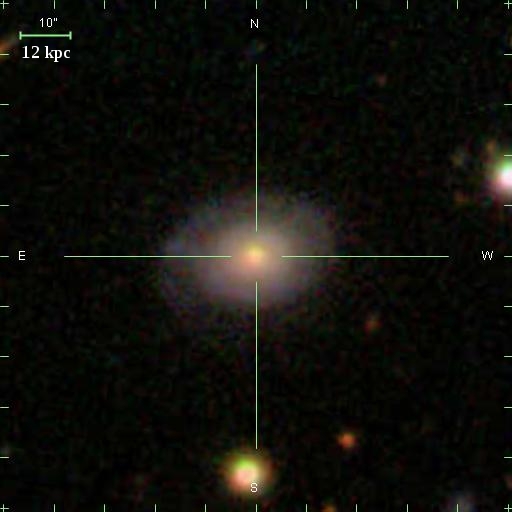}
\includegraphics[width=0.19\textwidth]{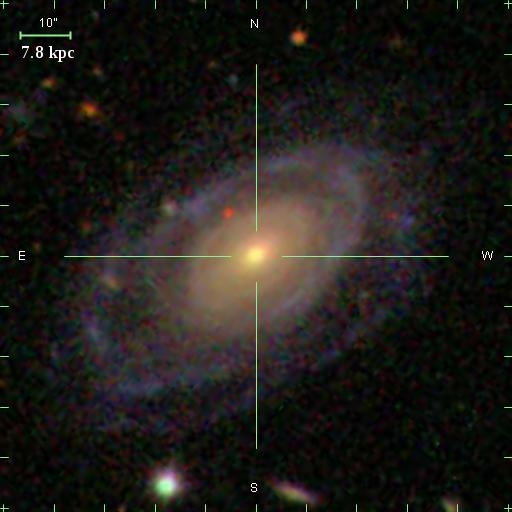}\\
\includegraphics[width=0.19\textwidth]{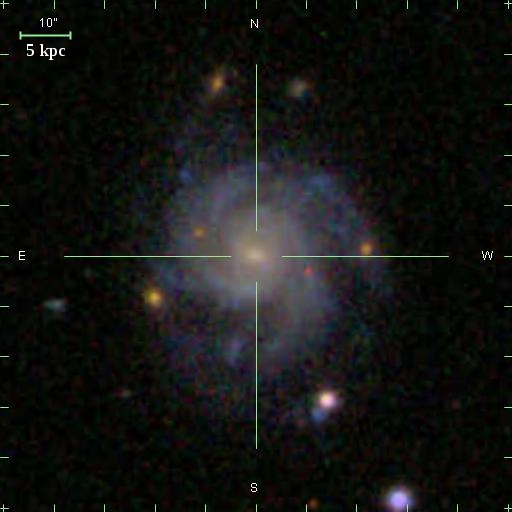}
\includegraphics[width=0.19\textwidth]{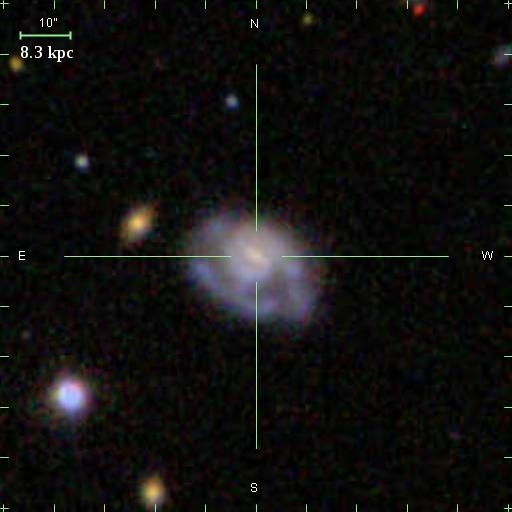}
\includegraphics[width=0.19\textwidth]{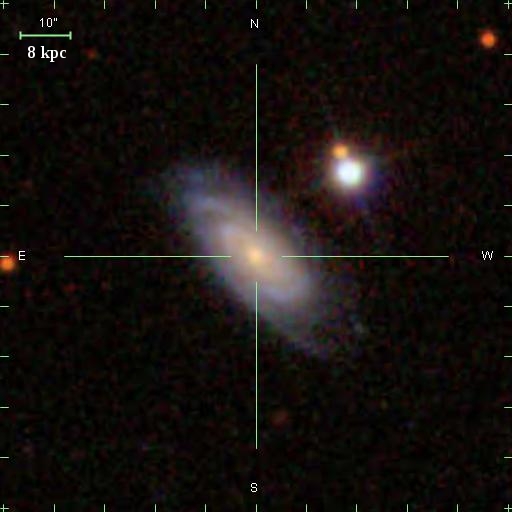}
\includegraphics[width=0.19\textwidth]{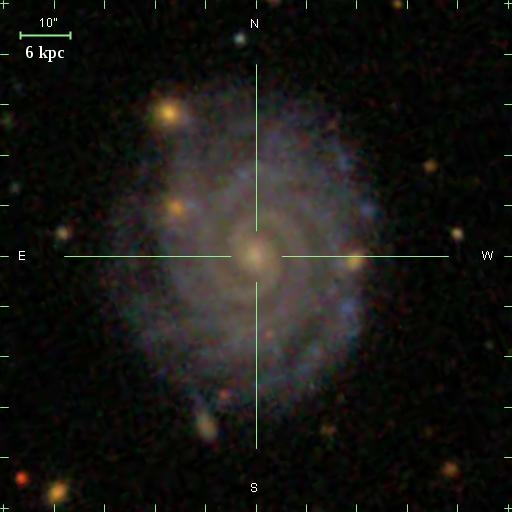}
\includegraphics[width=0.19\textwidth]{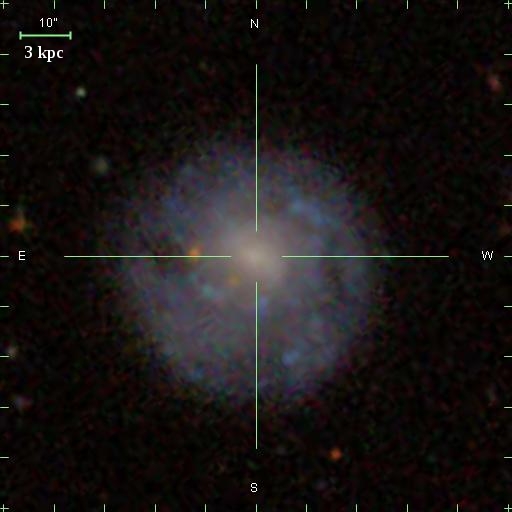}
\end{center}
\caption[The ten most isolated CIG galaxies in the SDSS-DR9 footprint]{SDSS three-colour images (field of view of 1\farcm7, North is up, East is to the left) of the ten most isolated CIG galaxies in the SDSS-DR9 footprint. From upper left to lower right: CIG 229, 245, 284, 318, 331, 541, 542, 546, 674, and 702.}\label{Fig:charts10}
\end{figure*}

Although only up to 15\% of the CIG galaxies in the sample have physically bound satellite, almost all galaxies (97\%) can be directly related to a LSS. The very large scatter in the quantification of the LSS (see the values spanned by $\eta_{k, \rm LSS}$ and $Q_{\rm LSS}$ in Fig.~\ref{Fig:isolparam}) shows that the CIG includes both galaxies dominated by their immediate environment as well as galaxies almost free from any external influence. In particular ten CIG galaxies are not associated to a LSS, at least within 3\,Mpc: CIG 229, 245, 284, 318, 331, 541, 542, 546, 674, and 702 (see their three-colour images in Fig.~\ref{Fig:charts10}).

The continuous distributions of the $\eta_{k, \rm LSS}$ and $Q_{\rm LSS}$ isolation parameters show that the CIG spans all the variety of environments between these two extreme cases. The connection of the CIG galaxies with the LSS is obvious due to the excess of similar redshift galaxies between 0.3 and 3\,Mpc, as can be seen in the lower panel of Fig.~\ref{Hist:diffvelold}. According to the middle and lower panels of Fig.~\ref{Fig:excess}, the large scale association is also noticeable due to higher number of brighter and bigger galaxies at redshift similar to those of the CIG galaxies, with respect to fainter, smaller background objects. Hence, the CIG galaxies are distributed following the LSS of the local Universe, although presenting a large heterogeneity in their degree of connection with it.

Some illustrations of the different environments around the CIG galaxies are shown in Fig.~\ref{Fig:charts}. Due to the large and roughly equivalent number of blue- and redshifted neighbour galaxies within the projected 3\,Mpc, the galaxy CIG\,771 may reside in the outskirts of a poor cluster. The galaxy CIG\,401 seems to be located towards the edge of a LSS, such as a filament or a wall. On the other hand, the galaxy CIG\,203 appears only mildly in relation with a LSS since only two LSS neighbours can be found in its environment. Its isolation parameters are very low ($\eta_{k, \rm LSS} = -1.94$ and $Q_{\rm LSS} = -5.65$) and its spatial location could be towards a void part of the local Universe. This environment is closer to the one of the ten galaxies for which we find no relation with the LSS within the first 3\,Mpc. It is interesting to note that, out of the ten most isolated galaxies studied here, nine are clearly late type spirals showing symmetric morphologies, with no visible signs of interaction (see Fig.~\ref{Fig:charts10}). Some of these galaxies could represent the closest remains of a fossil spiral population.

\subsection{Influence of the environment on the evolution of the primary galaxies}

To delimit the role of the environment on the physical properties of the galaxies we compare, within the CIG, the most isolated galaxies to the galaxies with companions. For this comparison, we use median values since they are less sensitive to outliers. Uncertainties are given by the 95\% confidence interval of the median. The subsample of galaxies with companions encloses galaxies with at least one physically bound satellite in their vicinity ($k_{\rm sat}$ or $k_{\rm sat, sup}$ strictly positive). The subsample containing the most isolated CIG galaxies incorporates the galaxies presenting the lowest values of the projected number density ($\eta_{k, \rm LSS} < -1.5$) and tidal strength ($Q_{\rm LSS} < -6$), along with the ten galaxies isolated from both their local and LSS environments (CIG 229, 245, 284, 318, 331, 541, 542, 546, 674, and 702). 

\begin{figure*}
\begin{center}
\includegraphics[width=\textwidth]{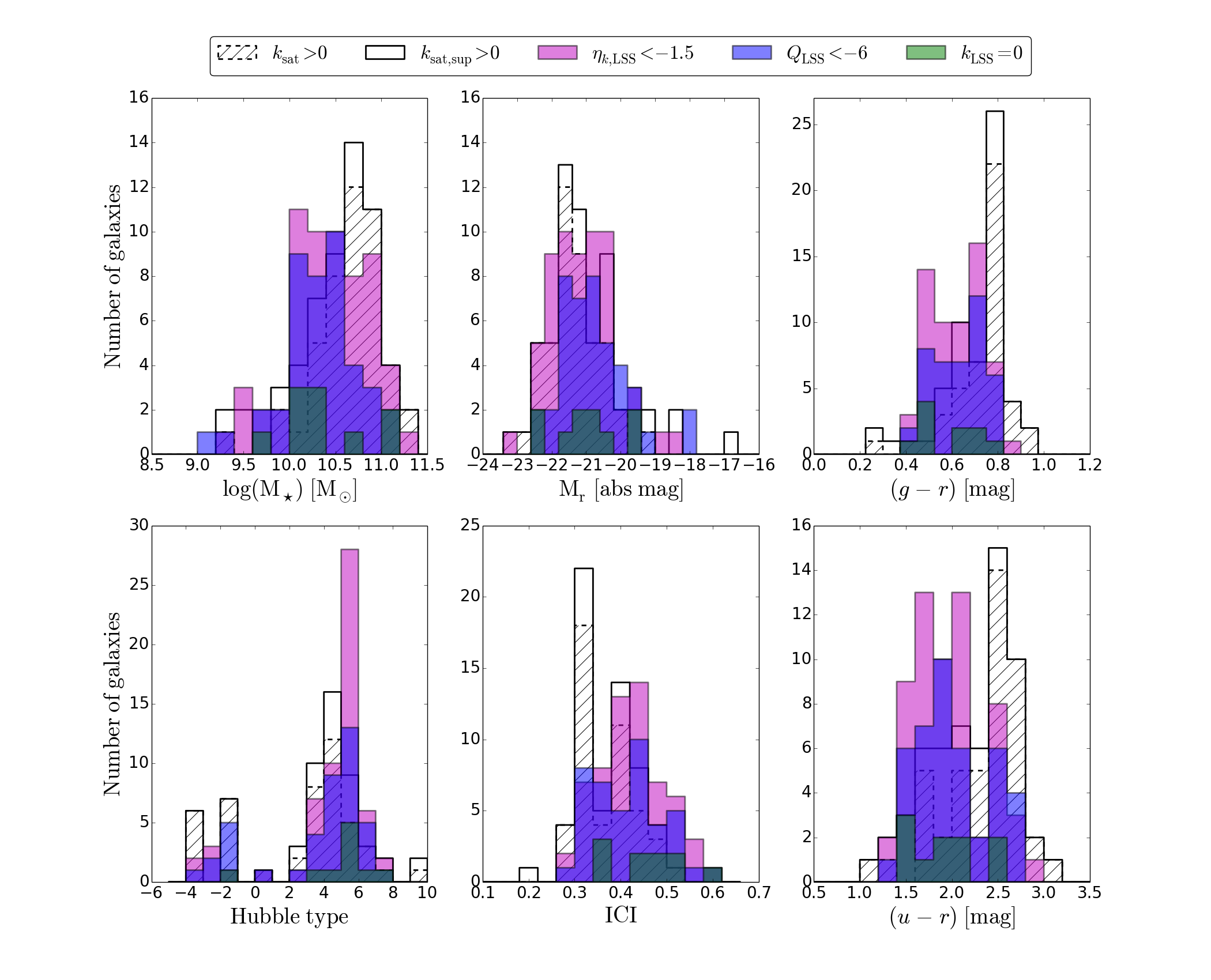}
\end{center}
\caption[CIG galaxy properties as a function of the environment]{Distributions of the stellar mass (upper left panel), r-band absolute magnitude (upper central panel), $(g-r)$ rest-frame colour (upper right panel), morphological Hubble type ($T$) according to \citet{2012A&A...540A..47F} (lower left panel), inverse concentration index (lower central panel), and $(u-r)$ rest-frame colour (lower right panel) for CIG galaxies. The distributions for the 10 most isolated CIG galaxies are represented by green histograms, and the most isolated CIG galaxies in terms of projected density $\eta_{k,\rm{LSS}}$ and tidal strength $Q_{\rm{LSS}}$ are represented by magenta and blue histograms, respectively. On the contrary, distributions for CIG galaxies with satellites are represented by white histograms (hatched histograms for the escape speed selection and plain white histograms in case of upper limit selection of satellites).} \label{Fig:hist_CIG_type_ICI_ur_mass_Mr_gr_1}
\end{figure*} 

In the upper left panel of Fig.~\ref{Fig:hist_CIG_type_ICI_ur_mass_Mr_gr_1}, the distribution of the logarithm of the stellar masses of the galaxies with companions (with median 10.70~$\pm$~0.10) and isolated galaxies (with median 10.35~$\pm$~0.17) are shown. Both subsamples extend from $10^{9}$ to $10^{11.5}$\,M$_\odot$, although the galaxies with companions might have a mild tendency (1$\sigma$) to be more massive, which may indicate a higher frequency of having suffered a merger in the past. Regarding the absolute magnitudes, the median for galaxies with companions is $-$21.48~$\pm$~0.22 , while for isolated galaxies these values are $-$20.98~$\pm$~0.27. There is no significant difference in the distributions of the absolute magnitudes between the two subsamples (upper central panel of Fig.~\ref{Fig:hist_CIG_type_ICI_ur_mass_Mr_gr_1}).

Rest-frame colours $(g-r)$ and $(u-r)$ are derived from the absolute magnitudes in $u$-, $g$- and $r$-band, where $u = M_u$, $g = M_g$, and $r = M_r$ magnitudes are corrected for Galactic extinction \citep[following the extinction maps from][]{1998ApJ...500..525S} and k-correction. The median $(g-r)$ value for the galaxies with physically bound satellites is 0.76~$\pm$~0.02, while the median value for the galaxies least affected by external tidal forces drops to 0.62~$\pm$~0.05 (upper right panel of Fig.~\ref{Fig:hist_CIG_type_ICI_ur_mass_Mr_gr_1}). This suggests that CIG galaxies with companions are in general redder and with older stellar populations with respect to the most isolated galaxies in the CIG which are bluer due to younger stellar populations (at a $\sim 2 \sigma$ confidence level).

In the lower left panel of Fig.~\ref{Fig:hist_CIG_type_ICI_ur_mass_Mr_gr_1}, the morphological $T$-types \citep{2012A&A...540A..47F} of the galaxies with companions and most isolated galaxies are shown. The fraction of early-type galaxies ($T$~$<$~0) is dominated by galaxies with companions. On the other hand, the most isolated galaxies concentrate around the Sc type ($T$~=~5), meaning that they are mainly late-type spiral galaxies. This trend is confirmed by the 10 most isolated galaxies: only one is early-type, while the remaining nine are consistently distributed around the Sc type. The inverse concentration index (ICI) defined as the ratio of the radii containing 50\% and 90\% of the Petrosian fluxes in the SDSS $r$-band, $C \equiv r_{p, 50} / r_{p, 90}$, is also an indicator of the morphological type. Early-type galaxies with a de Vaucouleurs profile will display values of the ICI around 0.3 while morphologies dominated by an exponential disk will show typical ICI values towards 0.43 \citep{2001AJ....122.1861S}. The ICI histograms in the lower central panel of Fig.~\ref{Fig:hist_CIG_type_ICI_ur_mass_Mr_gr_1} confirm the trends based on the visual (optical) morphology: a marginal ($1 \sigma$) segregation between early-type galaxies with companions (median 0.36~$\pm$~0.02) and isolated late-type galaxies (median 0.42~$\pm$~0.03). The intrinsic dispersion due to the use of the ICI as an estimation of the morphological type may also mix the two populations, which may be genuinely more separated. These tendencies can be related to the well known morphology-density relation for field and cluster galaxies \citep{1980ApJ...236..351D,1997ApJ...490..577D}, but it is noteworthy to appreciate it even when the local environment is defined by only one, two, or three faint satellites.

The stellar populations of primary galaxies can be characterised in terms of $(u-r)$ rest-frame colours. In the lower right panel of Fig.~\ref{Fig:hist_CIG_type_ICI_ur_mass_Mr_gr_1}, the well known SDSS-discovered bimodality appears, with an optimal separation at $(u-r)~=~2.22$ \citep{2001AJ....122.1861S}. Isolated galaxies mainly distribute in the range $(u-r)~<~2.22$ (median 1.95~$\pm$~0.16), while galaxies with companions spread over the area defined by $(u-r)~>~2.22$ (median 2.48~$\pm$~0.11). This segregation (with more than a 68\% level of confidence) means that isolated galaxies are in general bluer, with a younger stellar population and rather high star formation with respect to older, redder galaxies with companions. These colours, in combination with morphological trends previously noticed and the $(g-r)$ colours, lead to a coherent view where isolated star forming galaxies are separated from older early-type galaxies with companions.

\begin{figure}
\begin{center}
\includegraphics[width=\columnwidth]{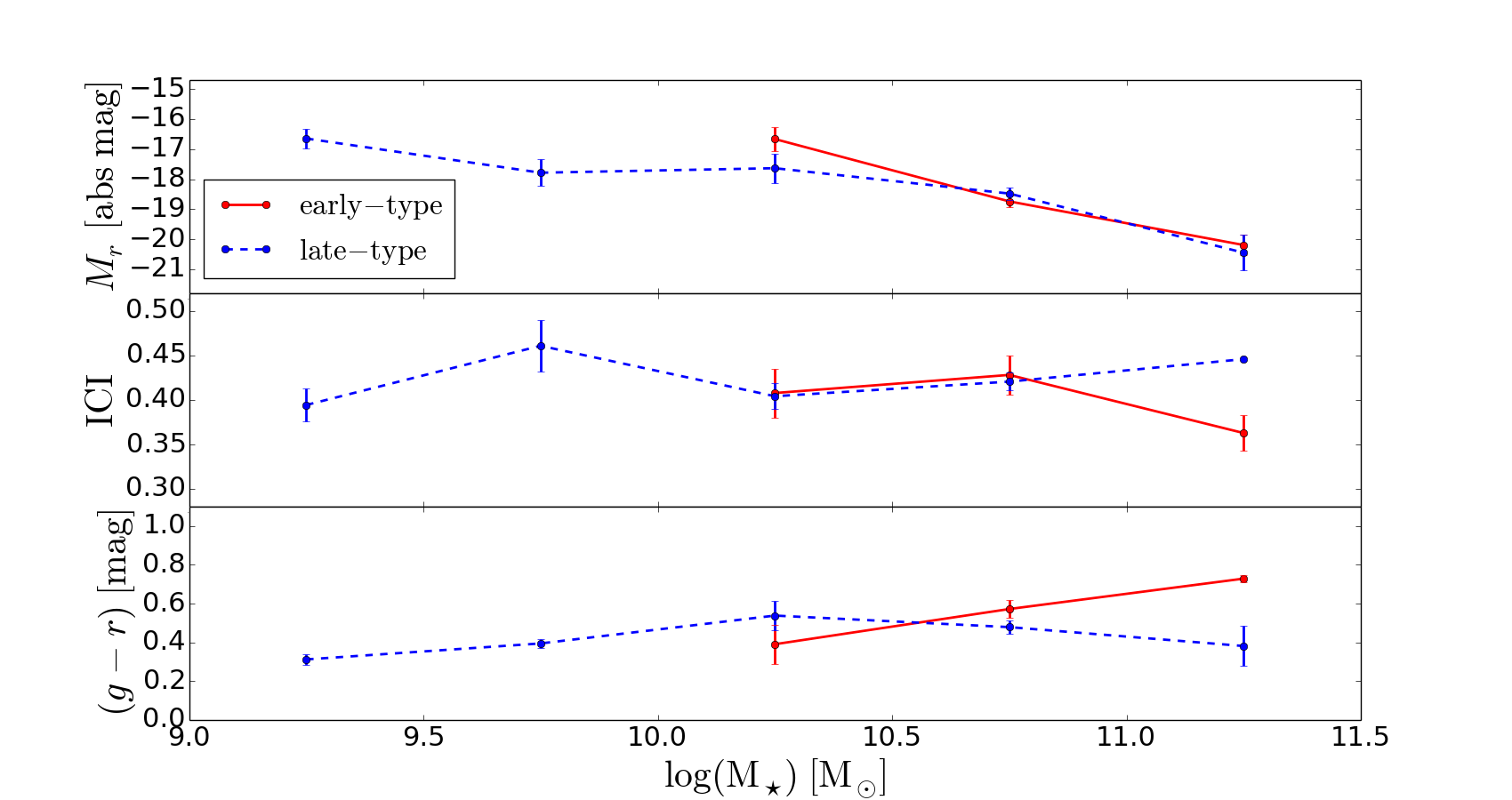}
\end{center}
\caption[Properties of satellites as a function of primary stellar mass]{Mean $r$-band absolute magnitude (upper panel), ICI (middle panel), and $(g-r)$ rest-frame colour (lower panel) for bound physical satellites as a function of CIG galaxy stellar mass. Mean satellite properties for early-type ($T$~$\leq$~0, red solid line) and late-type ($T$~$>$~0, blue dashed line) CIG galaxies are represented. Error bars are given by the standard deviation.} \label{Fig:sat_Mr_CI_color}
\end{figure}

In Fig.~\ref{Fig:sat_Mr_CI_color}, we show some mean properties of the physically linked satellites as a function of the stellar masses of the primary CIG galaxies, for two subsamples: physical satellites around early- and late-type CIG galaxies. The distribution of the absolute magnitudes of the satellites is comparable for both subsamples, although the most massive early-type CIG galaxies ($\rm M_{\star} > 10^{10.5}\,M_{\sun}$) may very marginally attract brighter satellites (see upper panel of Fig.~\ref{Fig:sat_Mr_CI_color}). A clearer tendency appears for the distribution of the ICI: massive early-type CIG galaxies will preferentially be surrounded by more early-type companions, with respect to late-type CIG galaxies which will present a higher fraction of late-type satellites (see central panel of Fig.~\ref{Fig:sat_Mr_CI_color}). This dichotomy is also seen in the $(g-r)$ colours of the satellites: the satellites are redder, likely have older stellar populations, around massive early-type CIG galaxies while they may present a younger stellar content around massive late-type CIG galaxies, as seen in the lower panel of Fig.~\ref{Fig:sat_Mr_CI_color}. 

This means that if the local environment has an influence on the evolution of the CIG galaxies, reciprocally, the satellites around the CIG galaxies may also be affected by the nature of the primary galaxy. This suggests that the CIG is composed by an heterogeneous population of galaxies, sampling old systems of galaxies but also spanning more recent, dynamical systems of galaxies.

%______________________________________________________________

\section{Summary and conclusions} \label{Sec:con}
   
We present a study of the 3-dimensional environment for a sample of 386 galaxies in the \textbf{C}atalogue of \textbf{I}solated \textbf{G}alaxies \citep[CIG;][]{1973AISAO...8....3K}, using the Ninth Data Release of the Sloan Digital Sky Survey (SDSS-DR9).

We identify and quantify the effects of the satellite distribution around a sample of galaxies in the CIG. To recover the physical satellites around the CIG galaxies, we first focus on the satellites which are within the escape speed of each CIG galaxy. We also propose a more conservative method based on the stacked Gaussian distribution of the velocity difference of the neighbours, which gives an upper limit to the influence of the local environment.

In comparison to a previous study \citep{2013A&A...560A...9A} we can estimate the effect of the physical associations that were not taken into account by the CIG isolation criteria, which could also have a non negligible influence on the evolution of the central CIG galaxy. The tidal strengths affecting the primary galaxy are estimated to quantify the effects of the local and Large Scale Structure (LSS) environments. To characterise the LSS around the CIG galaxies, we define the projected number density parameter at the 5$^{\rm th}$ nearest neighbour.

Our main conclusions are the following: 

\begin{enumerate}
\item Out of the 386 CIG galaxies considered in this study, at least 340 (88\% of the sample) have no physically linked satellite. Following the more conservative Gaussian distribution method to identify physical satellites around the CIG galaxies leads to upper limits: out of the 386 CIG galaxies, 327 galaxies (85\% of the sample) would have no physical companion within a projected distance of 0.3\,Mpc.

\item Consequently, about 12\% (and up to 15\%) of the CIG galaxies have physically bound satellites. CIG galaxies with companions might have a mild tendency (0.3-0.4 dex) to be more massive, which suggests a higher frequency of having suffered a merger in the past. Satellites are in general redder, brighter, and bigger for more massive central CIG galaxies. Also, massive elliptical and lenticular CIG galaxies tend to have satellites with earlier types than similar mass spiral CIG galaxies.

\item Although 15\% at most of the CIG galaxies in the sample have physically bound satellite, almost all galaxies (97\%) can be directly related to a LSS. The very large scatter in the quantification of the LSS shows that the CIG includes both galaxies dominated by their immediate environment as well as galaxies almost free from any external influence.

\item The continuous distributions of the $\eta_{k, \rm LSS}$ and $Q_{\rm LSS}$ isolation parameters show that the CIG spans a variety of environments. The connection of the CIG galaxies with the LSS is obvious due to the excess of similar redshift galaxies between 0.3 and 3\,Mpc. The CIG galaxies are distributed following the LSS of the local universe, although presenting a large heterogeneity in their degrees of connection with it.

\item To evaluate the role of the physically bound satellites with respect to the large scale environment, we compare the magnitudes of the sum of the tidal strengths produced by the physical companions to the sum of the tidal strengths created by all the galaxies in the LSS. When at least one physical companion is present near a CIG galaxy, its effect largely dominates (usually more than 90\%) the tidal strengths generated by the LSS.

\item To delimit the role of the environment on the physical properties of the galaxies we compare, within the CIG, the most isolated galaxies to the galaxies with companions. We find a clear segregation between CIG galaxies with companions and isolated CIG galaxies. Isolated galaxies are in general bluer, with a younger stellar population and rather high star formation with respect to the older, redder galaxies with companions. These $(u-r)$ colours, in combination with the morphological trends and the $(g-r)$ colours, lead to a coherent view where isolated star forming galaxies are separated from older elliptical galaxies with companions.

\item Conjointly, we find that the satellites are redder and with older stellar populations around massive early-type CIG galaxies while they have a younger stellar content around massive late-type CIG galaxies. This means that if the local environment has an influence on the evolution of the CIG galaxies, reciprocally, the satellites around the CIG galaxies may also be affected by the nature of the primary galaxy. This suggests that the CIG is composed of a heterogeneous population of galaxies, sampling old systems of galaxies but also spanning more recent, dynamical systems of galaxies.

\item As mentioned, the CIG samples a variety of environments, from galaxies in interaction with physical satellites to galaxies with no neighbours in the first 3\,Mpc around them. Hence, in the construction of catalogues of galaxies in relation to their environments (isolated, pairs, triplets, groups of galaxies), redshift surveys are required in order to distinguish small, faint, physically bound satellites from a background projected galaxy population and reach a more comprehensive 3-dimensional picture of the surroundings.

\end{enumerate}

%______________________________________________________________

\begin{acknowledgements}
The authors acknowledge the referee for his/her very detailed and useful report, which helped to clarify and improve the quality of this work.

The authors thank J. Jim\'enez-Vicente for useful discussions.

This work has been supported by Grant AYA2011-30491-C02-01, co-financed by MICINN and FEDER funds, and the Junta de Andalucía (Spain) grants P08-FQM-4205 and TIC-114, as well as under the EU 7th Framework Programme in the area of Digital Libraries and Digital Preservation. (ICT-2009.4.1) Project reference: 270192.

This work was partially supported by a Junta de Andalucía Grant FQM108 and a Spanish MEC Grant AYA-2007-67625-C02-02.

Funding for SDSS-III has been provided by the Alfred P. Sloan Foundation, the Participating Institutions, the National Science Foundation, and the U.S. Department of Energy Office of Science. The SDSS-III web site is http://www.sdss3.org/. SDSS-III is managed by the Astrophysical Research Consortium for the Participating Institutions of the SDSS-III Collaboration including the University of Arizona, the Brazilian Participation Group, Brookhaven National Laboratory, University of Cambridge, University of Florida, the French Participation Group, the German Participation Group, the Instituto de Astrofisica de Canarias, the Michigan State/Notre Dame/JINA Participation Group, Johns Hopkins University, Lawrence Berkeley National Laboratory, Max Planck Institute for Astrophysics, New Mexico State University, New York University, Ohio State University, Pennsylvania State University, University of Portsmouth, Princeton University, the Spanish Participation Group, University of Tokyo, University of Utah, Vanderbilt University, University of Virginia, University of Washington, and Yale University.

This research has made use of data obtained using, or software provided by, the UK's AstroGrid Virtual Observatory Project, which is funded by the Science and Technology Facilities Council and through the EU's Framework 6 programme. We also acknowledge the use of STILTS and TOPCAT tools \citet{2005ASPC..347...29T}.

This research made use of python ({\tt http://www.python.org}), of Matplotlib \citep{Hunter:2007}, a suite of open-source python modules that provides a framework for creating scientific plots.

This research has made use of the NASA/IPAC Extragalactic Database (NED) which is operated by the Jet Propulsion Laboratory, California Institute of Technology, under contract with the National Aeronautics and Space Administration. 

We acknowledge the usage of the HyperLeda database ({\tt http://leda.univ-lyon1.fr}) \citep{2003A&A...412...45P}.

We thank the SAO/NASA Astrophysics Data System (ADS) that is always so useful.
\end{acknowledgements}

\bibliography{astroph}

\begin{thebibliography}{43}
\expandafter\ifx\csname natexlab\endcsname\relax\def\natexlab#1{#1}\fi

\bibitem[{{Agustsson} \& {Brainerd}(2010)}]{2010ApJ...709.1321A}
{Agustsson}, I. \& {Brainerd}, T.~G. 2010, \apj, 709, 1321

\bibitem[{{Ahn} {et~al.}(2012){Ahn}, {Alexandroff}, {Allende Prieto},
  {Anderson}, {Anderton}, {Andrews}, {Aubourg}, {Bailey}, {Balbinot}, {Barnes},
  \& et~al.}]{2012ApJS..203...21A}
{Ahn}, C.~P., {Alexandroff}, R., {Allende Prieto}, C., {et~al.} 2012, \apjs,
  203, 21

\bibitem[{{Anderhalden} {et~al.}(2013){Anderhalden}, {Schneider}, {Macci{\`o}},
  {Diemand}, \& {Bertone}}]{2013JCAP...03..014A}
{Anderhalden}, D., {Schneider}, A., {Macci{\`o}}, A.~V., {Diemand}, J., \&
  {Bertone}, G. 2013, \jcap, 3, 14

\bibitem[{{Argudo-Fern{\'a}ndez} {et~al.}(2013){Argudo-Fern{\'a}ndez},
  {Verley}, {Bergond}, {Sulentic}, {Sabater}, {Fern{\'a}ndez Lorenzo}, {Leon},
  {Espada}, {Verdes-Montenegro}, {Santander-Vela}, {Ruiz}, \&
  {S{\'a}nchez-Exp{\'o}sito}}]{2013A&A...560A...9A}
{Argudo-Fern{\'a}ndez}, M., {Verley}, S., {Bergond}, G., {et~al.} 2013, \aap,
  560, A9

\bibitem[{{Blanton} \& {Roweis}(2007)}]{2007AJ....133..734B}
{Blanton}, M.~R. \& {Roweis}, S. 2007, \aj, 133, 734

\bibitem[{{Bozek} {et~al.}(2013){Bozek}, {Wyse}, \&
  {Gilmore}}]{2013ApJ...772..109B}
{Bozek}, B., {Wyse}, R.~F.~G., \& {Gilmore}, G. 2013, \apj, 772, 109

\bibitem[{{Choi} {et~al.}(2007){Choi}, {Weinberg}, \&
  {Katz}}]{2007AAS...21112602C}
{Choi}, J.-H., {Weinberg}, M.~D., \& {Katz}, N. 2007, in Bulletin of the
  American Astronomical Society, Vol.~39, American Astronomical Society Meeting
  Abstracts, 126.02

\bibitem[{{Cooper} {et~al.}(2012){Cooper}, {Griffith}, {Newman}, {Coil},
  {Davis}, {Dutton}, {Faber}, {Guhathakurta}, {Koo}, {Lotz}, {Weiner},
  {Willmer}, \& {Yan}}]{2012MNRAS.419.3018C}
{Cooper}, M.~C., {Griffith}, R.~L., {Newman}, J.~A., {et~al.} 2012, \mnras,
  419, 3018

\bibitem[{{Dawson} {et~al.}(2013){Dawson}, {Schlegel}, {Ahn}, {Anderson},
  {Aubourg}, {Bailey}, {Barkhouser}, {Bautista}, {Beifiori}, {Berlind},
  {Bhardwaj}, {Bizyaev}, {Blake}, {Blanton}, {Blomqvist}, {Bolton}, {Borde},
  {Bovy}, {Brandt}, {Brewington}, {Brinkmann}, {Brown}, {Brownstein}, {Bundy},
  {Busca}, {Carithers}, {Carnero}, {Carr}, {Chen}, {Comparat}, {Connolly},
  {Cope}, {Croft}, {Cuesta}, {da Costa}, {Davenport}, {Delubac}, {de Putter},
  {Dhital}, {Ealet}, {Ebelke}, {Eisenstein}, {Escoffier}, {Fan}, {Filiz Ak},
  {Finley}, {Font-Ribera}, {G{\'e}nova-Santos}, {Gunn}, {Guo}, {Haggard},
  {Hall}, {Hamilton}, {Harris}, {Harris}, {Ho}, {Hogg}, {Holder}, {Honscheid},
  {Huehnerhoff}, {Jordan}, {Jordan}, {Kauffmann}, {Kazin}, {Kirkby}, {Klaene},
  {Kneib}, {Le Goff}, {Lee}, {Long}, {Loomis}, {Lundgren}, {Lupton}, {Maia},
  {Makler}, {Malanushenko}, {Malanushenko}, {Mandelbaum}, {Manera}, {Maraston},
  {Margala}, {Masters}, {McBride}, {McDonald}, {McGreer}, {McMahon}, {Mena},
  {Miralda-Escud{\'e}}, {Montero-Dorta}, {Montesano}, {Muna}, {Myers},
  {Naugle}, {Nichol}, {Noterdaeme}, {Nuza}, {Olmstead}, {Oravetz}, {Oravetz},
  {Owen}, {Padmanabhan}, {Palanque-Delabrouille}, {Pan}, {Parejko},
  {P{\^a}ris}, {Percival}, {P{\'e}rez-Fournon}, {P{\'e}rez-R{\`a}fols},
  {Petitjean}, {Pfaffenberger}, {Pforr}, {Pieri}, {Prada}, {Price-Whelan},
  {Raddick}, {Rebolo}, {Rich}, {Richards}, {Rockosi}, {Roe}, {Ross}, {Ross},
  {Rossi}, {Rubi{\~n}o-Martin}, {Samushia}, {S{\'a}nchez}, {Sayres}, {Schmidt},
  {Schneider}, {Sc{\'o}ccola}, {Seo}, {Shelden}, {Sheldon}, {Shen}, {Shu},
  {Slosar}, {Smee}, {Snedden}, {Stauffer}, {Steele}, {Strauss}, {Streblyanska},
  {Suzuki}, {Swanson}, {Tal}, {Tanaka}, {Thomas}, {Tinker}, {Tojeiro},
  {Tremonti}, {Vargas Maga{\~n}a}, {Verde}, {Viel}, {Wake}, {Watson}, {Weaver},
  {Weinberg}, {Weiner}, {West}, {White}, {Wood-Vasey}, {Yeche}, {Zehavi},
  {Zhao}, \& {Zheng}}]{2013AJ....145...10D}
{Dawson}, K.~S., {Schlegel}, D.~J., {Ahn}, C.~P., {et~al.} 2013, \aj, 145, 10

\bibitem[{{Diaferio} \& {Geller}(1997)}]{1997ApJ...481..633D}
{Diaferio}, A. \& {Geller}, M.~J. 1997, \apj, 481, 633

\bibitem[{{Dressler}(1980)}]{1980ApJ...236..351D}
{Dressler}, A. 1980, \apj, 236, 351

\bibitem[{{Dressler} {et~al.}(1997){Dressler}, {Oemler}, {Couch}, {Smail},
  {Ellis}, {Barger}, {Butcher}, {Poggianti}, \&
  {Sharples}}]{1997ApJ...490..577D}
{Dressler}, A., {Oemler}, Jr., A., {Couch}, W.~J., {et~al.} 1997, \apj, 490,
  577

\bibitem[{{Edman} {et~al.}(2012){Edman}, {Barton}, \&
  {Bullock}}]{2012MNRAS.424.1454E}
{Edman}, J.~P., {Barton}, E.~J., \& {Bullock}, J.~S. 2012, \mnras, 424, 1454

\bibitem[{{Einasto} \& {Einasto}(1987)}]{1987MNRAS.226..543E}
{Einasto}, M. \& {Einasto}, J. 1987, \mnras, 226, 543

\bibitem[{{Eisenstein} {et~al.}(2011){Eisenstein}, {Weinberg}, {Agol},
  {Aihara}, {Allende Prieto}, {Anderson}, {Arns}, {Aubourg}, {Bailey},
  {Balbinot}, \& et~al.}]{2011AJ....142...72E}
{Eisenstein}, D.~J., {Weinberg}, D.~H., {Agol}, E., {et~al.} 2011, \aj, 142, 72

\bibitem[{{Fern{\'a}ndez Lorenzo} {et~al.}(2013){Fern{\'a}ndez Lorenzo},
  {Sulentic}, {Verdes-Montenegro}, \&
  {Argudo-Fern{\'a}ndez}}]{2013MNRAS.434..325F}
{Fern{\'a}ndez Lorenzo}, M., {Sulentic}, J., {Verdes-Montenegro}, L., \&
  {Argudo-Fern{\'a}ndez}, M. 2013, \mnras, 434, 325

\bibitem[{{Fern{\'a}ndez Lorenzo} {et~al.}(2012){Fern{\'a}ndez Lorenzo},
  {Sulentic}, {Verdes-Montenegro}, {Ruiz}, {Sabater}, \&
  {S{\'a}nchez}}]{2012A&A...540A..47F}
{Fern{\'a}ndez Lorenzo}, M., {Sulentic}, J., {Verdes-Montenegro}, L., {et~al.}
  2012, \aap, 540, A47

\bibitem[{{Ferrero} {et~al.}(2012){Ferrero}, {Abadi}, {Navarro}, {Sales}, \&
  {Gurovich}}]{2012MNRAS.425.2817F}
{Ferrero}, I., {Abadi}, M.~G., {Navarro}, J.~F., {Sales}, L.~V., \& {Gurovich},
  S. 2012, \mnras, 425, 2817

\bibitem[{{Gonz{\'a}lez} {et~al.}(2013){Gonz{\'a}lez}, {Kravtsov}, \&
  {Gnedin}}]{2013ApJ...770...96G}
{Gonz{\'a}lez}, R.~E., {Kravtsov}, A.~V., \& {Gnedin}, N.~Y. 2013, \apj, 770,
  96

\bibitem[{{Guo} {et~al.}(2011){Guo}, {Cole}, {Eke}, \&
  {Frenk}}]{2011MNRAS.417..370G}
{Guo}, Q., {Cole}, S., {Eke}, V., \& {Frenk}, C. 2011, \mnras, 417, 370

\bibitem[{{Guo} {et~al.}(2012){Guo}, {Cole}, {Eke}, \&
  {Frenk}}]{2012MNRAS.427..428G}
{Guo}, Q., {Cole}, S., {Eke}, V., \& {Frenk}, C. 2012, \mnras, 427, 428

\bibitem[{Hunter(2007)}]{Hunter:2007}
Hunter, J.~D. 2007, Computing In Science \& Engineering, 9, 90

\bibitem[{{Kaiser}(1987)}]{1987MNRAS.227....1K}
{Kaiser}, N. 1987, \mnras, 227, 1

\bibitem[{{Karachentseva}(1973)}]{1973AISAO...8....3K}
{Karachentseva}, V.~E. 1973, Astrofizicheskie Issledovaniia Izvestiya
  Spetsial'noj Astrofizicheskoj Observatorii, 8, 3

\bibitem[{{Karachentseva} {et~al.}(2011){Karachentseva}, {Karachentsev}, \&
  {Melnyk}}]{2011AstBu..66..389K}
{Karachentseva}, V.~E., {Karachentsev}, I.~D., \& {Melnyk}, O.~V. 2011,
  Astrophysical Bulletin, 66, 389

\bibitem[{{Moster} {et~al.}(2013){Moster}, {Naab}, \&
  {White}}]{2013MNRAS.428.3121M}
{Moster}, B.~P., {Naab}, T., \& {White}, S.~D.~M. 2013, \mnras, 428, 3121

\bibitem[{{Nair} \& {Abraham}(2010)}]{2010ApJS..186..427N}
{Nair}, P.~B. \& {Abraham}, R.~G. 2010, \apjs, 186, 427

\bibitem[{{Park} {et~al.}(2007){Park}, {Choi}, {Vogeley}, {Gott}, \&
  {Blanton}}]{2007ApJ...658..898P}
{Park}, C., {Choi}, Y., {Vogeley}, M.~S., {Gott}, I. J.~R., \& {Blanton}, M.~R.
  2007, \apj, 658, 898

\bibitem[{{Paturel} {et~al.}(2003){Paturel}, {Petit}, {Prugniel}, {Theureau},
  {Rousseau}, {Brouty}, {Dubois}, \& {Cambr{\'e}sy}}]{2003A&A...412...45P}
{Paturel}, G., {Petit}, C., {Prugniel}, P., {et~al.} 2003, \aap, 412, 45

\bibitem[{{Prada} {et~al.}(2003){Prada}, {Vitvitska}, {Klypin}, {Holtzman},
  {Schlegel}, {Grebel}, {Rix}, {Brinkmann}, {McKay}, \&
  {Csabai}}]{2003ApJ...598..260P}
{Prada}, F., {Vitvitska}, M., {Klypin}, A., {et~al.} 2003, \apj, 598, 260

\bibitem[{{Sabater} {et~al.}(2013){Sabater}, {Best}, \&
  {Argudo-Fern{\'a}ndez}}]{2013MNRAS.430..638S}
{Sabater}, J., {Best}, P.~N., \& {Argudo-Fern{\'a}ndez}, M. 2013, \mnras, 430,
  638

\bibitem[{{Sales} \& {Lambas}(2005)}]{2005MNRAS.356.1045S}
{Sales}, L. \& {Lambas}, D.~G. 2005, \mnras, 356, 1045

\bibitem[{{Schlegel} {et~al.}(1998){Schlegel}, {Finkbeiner}, \&
  {Davis}}]{1998ApJ...500..525S}
{Schlegel}, D.~J., {Finkbeiner}, D.~P., \& {Davis}, M. 1998, \apj, 500, 525

\bibitem[{{Smee} {et~al.}(2012){Smee}, {Gunn}, {Uomoto}, {Roe}, {Schlegel},
  {Rockosi}, {Carr}, {Leger}, {Dawson}, {Olmstead}, {Brinkmann}, {Owen},
  {Barkhouser}, {Honscheid}, {Harding}, {Long}, {Lupton}, {Loomis}, {Anderson},
  {Annis}, {Bernardi}, {Bhardwaj}, {Bizyaev}, {Bolton}, {Brewington}, {Briggs},
  {Burles}, {Burns}, {Castander}, {Connolly}, {Davenport}, {Ebelke}, {Epps},
  {Feldman}, {Friedman}, {Frieman}, {Heckman}, {Hull}, {Knapp}, {Lawrence},
  {Loveday}, {Mannery}, {Malanushenko}, {Malanushenko}, {Merrelli}, {Muna},
  {Newman}, {Nichol}, {Oravetz}, {Pan}, {Pope}, {Ricketts}, {Shelden},
  {Sandford}, {Siegmund}, {Simmons}, {Smith}, {Snedden}, {Schneider},
  {Strauss}, {SubbaRao}, {Tremonti}, {Waddell}, \&
  {York}}]{2012arXiv1208.2233S}
{Smee}, S., {Gunn}, J.~E., {Uomoto}, A., {et~al.} 2012, ArXiv e-prints

\bibitem[{{Strateva} {et~al.}(2001){Strateva}, {Ivezi{\'c}}, {Knapp},
  {Narayanan}, {Strauss}, {Gunn}, {Lupton}, {Schlegel}, {Bahcall}, {Brinkmann},
  {Brunner}, {Budav{\'a}ri}, {Csabai}, {Castander}, {Doi}, {Fukugita}, {Gy{\H
  o}ry}, {Hamabe}, {Hennessy}, {Ichikawa}, {Kunszt}, {Lamb}, {McKay},
  {Okamura}, {Racusin}, {Sekiguchi}, {Schneider}, {Shimasaku}, \&
  {York}}]{2001AJ....122.1861S}
{Strateva}, I., {Ivezi{\'c}}, {\v Z}., {Knapp}, G.~R., {et~al.} 2001, \aj, 122,
  1861

\bibitem[{{Strauss} {et~al.}(2002){Strauss}, {Weinberg}, {Lupton}, {Narayanan},
  {Annis}, {Bernardi}, {Blanton}, {Burles}, {Connolly}, {Dalcanton}, {Doi},
  {Eisenstein}, {Frieman}, {Fukugita}, {Gunn}, \v{Z}. {Ivezi{\'c}}, {Kent},
  {Kim}, {Knapp}, {Kron}, {Munn}, {Newberg}, {Nichol}, {Okamura}, {Quinn},
  {Richmond}, {Schlegel}, {Shimasaku}, {SubbaRao}, {Szalay}, {Vanden Berk},
  {Vogeley}, {Yanny}, {Yasuda}, {York}, \& {Zehavi}}]{2002AJ....124.1810S}
{Strauss}, M.~A., {Weinberg}, D.~H., {Lupton}, R.~H., {et~al.} 2002, \aj, 124,
  1810

\bibitem[{{Strauss} \& {Willick}(1995)}]{1995PhR...261..271S}
{Strauss}, M.~A. \& {Willick}, J.~A. 1995, \physrep, 261, 271

\bibitem[{{Sulentic} {et~al.}(2006){Sulentic}, {Verdes-Montenegro}, {Bergond},
  {Lisenfeld}, {Durbala}, {Espada}, {Garcia}, {Leon}, {Sabater}, {Verley},
  {Casanova}, \& {Sota}}]{2006A&A...449..937S}
{Sulentic}, J.~W., {Verdes-Montenegro}, L., {Bergond}, G., {et~al.} 2006, \aap,
  449, 937

\bibitem[{{Taylor}(2005)}]{2005ASPC..347...29T}
{Taylor}, M.~B. 2005, in Astronomical Society of the Pacific Conference Series,
  Vol. 347, Astronomical Data Analysis Software and Systems XIV, ed.
  {P.~Shopbell, M.~Britton, \& R.~Ebert}, 29

\bibitem[{{Verdes-Montenegro} {et~al.}(2005){Verdes-Montenegro}, {Sulentic},
  {Lisenfeld}, {Leon}, {Espada}, {Garcia}, {Sabater}, \&
  {Verley}}]{2005A&A...436..443V}
{Verdes-Montenegro}, L., {Sulentic}, J., {Lisenfeld}, U., {et~al.} 2005, \aap,
  436, 443

\bibitem[{{Verley} {et~al.}(2007{\natexlab{a}}){Verley}, {Leon},
  {Verdes-Montenegro}, {Combes}, {Sabater}, {Sulentic}, {Bergond}, {Espada},
  {Garc{\'i}a}, {Lisenfeld}, \& {Odewahn}}]{2007A&A...472..121V}
{Verley}, S., {Leon}, S., {Verdes-Montenegro}, L., {et~al.} 2007{\natexlab{a}},
  \aap, 472, 121

\bibitem[{{Verley} {et~al.}(2007{\natexlab{b}}){Verley}, {Odewahn},
  {Verdes-Montenegro}, {Leon}, {Combes}, {Sulentic}, {Bergond}, {Espada},
  {Garc{\'i}a}, {Lisenfeld}, \& {Sabater}}]{2007A&A...470..505V}
{Verley}, S., {Odewahn}, S.~C., {Verdes-Montenegro}, L., {et~al.}
  2007{\natexlab{b}}, \aap, 470, 505

\bibitem[{{York} {et~al.}(2000){York}, {Adelman}, {Anderson}, {Anderson},
  {Annis}, {Bahcall}, {Bakken}, {Barkhouser}, {Bastian}, {Berman}, {Boroski},
  {Bracker}, {Briegel}, {Briggs}, {Brinkmann}, {Brunner}, {Burles}, {Carey},
  {Carr}, {Castander}, {Chen}, {Colestock}, {Connolly}, {Crocker}, {Csabai},
  {Czarapata}, {Davis}, {Doi}, {Dombeck}, {Eisenstein}, {Ellman}, {Elms},
  {Evans}, {Fan}, {Federwitz}, {Fiscelli}, {Friedman}, {Frieman}, {Fukugita},
  {Gillespie}, {Gunn}, {Gurbani}, {de Haas}, {Haldeman}, {Harris}, {Hayes},
  {Heckman}, {Hennessy}, {Hindsley}, {Holm}, {Holmgren}, {Huang}, {Hull},
  {Husby}, {Ichikawa}, {Ichikawa}, \v{Z}. {Ivezi{\'c}}, {Kent}, {Kim},
  {Kinney}, {Klaene}, {Kleinman}, {Kleinman}, {Knapp}, {Korienek}, {Kron},
  {Kunszt}, {Lamb}, {Lee}, {Leger}, {Limmongkol}, {Lindenmeyer}, {Long},
  {Loomis}, {Loveday}, {Lucinio}, {Lupton}, {MacKinnon}, {Mannery}, {Mantsch},
  {Margon}, {McGehee}, {McKay}, {Meiksin}, {Merelli}, {Monet}, {Munn},
  {Narayanan}, {Nash}, {Neilsen}, {Neswold}, {Newberg}, {Nichol}, {Nicinski},
  {Nonino}, {Okada}, {Okamura}, {Ostriker}, {Owen}, {Pauls}, {Peoples},
  {Peterson}, {Petravick}, {Pier}, {Pope}, {Pordes}, {Prosapio},
  {Rechenmacher}, {Quinn}, {Richards}, {Richmond}, {Rivetta}, {Rockosi},
  {Ruthmansdorfer}, {Sandford}, {Schlegel}, {Schneider}, {Sekiguchi}, {Sergey},
  {Shimasaku}, {Siegmund}, {Smee}, {Smith}, {Snedden}, {Stone}, {Stoughton},
  {Strauss}, {Stubbs}, {SubbaRao}, {Szalay}, {Szapudi}, {Szokoly}, {Thakar},
  {Tremonti}, {Tucker}, {Uomoto}, {Vanden Berk}, {Vogeley}, {Waddell}, {Wang},
  {Watanabe}, {Weinberg}, {Yanny}, \& {Yasuda}}]{2000AJ....120.1579Y}
{York}, D.~G., {Adelman}, J., {Anderson}, J. J.~E., {et~al.} 2000, \aj, 120,
  1579

\end{thebibliography}

% \end{document}

\Online
% \begin{appendix}
\addtocounter{table}{1} 
\onllongtab{1}{
% \begin{landscape}
\begin{longtable}{cccccccccc}
\caption{\label{Tab:isolparam}Quantification of the environment.
The columns correspond to:
(1) galaxy identification according to the CIG;
(2) $Q_{\rm{sat}}$, tidal strength estimation for satellites using the escape speed method;
(3) $Q_{\rm{sat,sup}}$, tidal strength estimation for satellites using the Gaussian distribution method;
(4) $k_{\rm{sat}}$, number of satellites selected following the escape speed method;
(5) $k_{\rm{sat,sup}}$, number of satellites selected following the Gaussian distribution method;
(6) $Q_{\rm{LSS}}$, tidal strength estimation of the LSS;
(7) $\eta_{k,\rm{LSS}}$, local number density estimation of the LSS;
(8) $k_{\rm{LSS}}$, number of LSS associations;
(9) $\frac{Q_{\rm{sat}}}{Q_{\rm{LSS}}}$, relation between $Q_{\rm{sat}}$ and $Q_{\rm{LSS}}$, from 0 to 1 if the considered satellites amount from 0\% to 100\% to the total tidal strength;
(10) $\frac{Q_{\rm{sat,sup}}}{Q_{\rm{LSS}}}$, relation between $Q_{\rm{sat,sup}}$ and $Q_{\rm{LSS}}$, from 0 to 1 if the considered satellites amount from 0\% to 100\% to the total tidal strength.}\\
\hline
\hline
(1) & (2) & (3) & (4) & (5) & (6) & (7) & (8) & (9) & (10) \\
CIG   &   $Q_{\rm{sat}}$   &   $Q_{\rm{sat,sup}}$  &  $k_{\rm{sat}}$  &  $k_{\rm{sat,sup}}$  &  $Q_{\rm{LSS}}$   &   $\eta_{k,\rm{LSS}}$  &  $k_{\rm{LSS}}$  &  $\frac{Q_{\rm{sat}}}{Q_{\rm{LSS}}}$  &  $\frac{Q_{\rm{sat,sup}}}{Q_{\rm{LSS}}}$ \\
\hline
\endfirsthead
\caption{Continued.}\\
\hline
(1) & (2) & (3) & (4) & (5) & (6) & (7) & (8) & (9) & (10) \\
CIG   &   $Q_{\rm{sat}}$   &   $Q_{\rm{sat,sup}}$  &  $k_{\rm{sat}}$  &  $k_{\rm{sat,sup}}$  &  $Q_{\rm{LSS}}$   &   $\eta_{k,\rm{LSS}}$  &  $k_{\rm{LSS}}$  &  $\frac{Q_{\rm{sat}}}{Q_{\rm{LSS}}}$  &  $\frac{Q_{\rm{sat,sup}}}{Q_{\rm{LSS}}}$ \\
\hline
\endhead
\hline
\endfoot
%%
% #CIG   &   Q_sat   &   Q_sat_sup &  k_sat  &  k_sat_sup  &  Q_LSS   &  eta_k_LSS  &   k_LSS  &  Q_sat/Q_LSS  &  Q_sat_sup/Q_LSS  \\
   11     &     NULL     &     NULL     &     0     &     0     &    -6.62     &    -1.43     &     5     &         NULL     &         NULL    \\
   33     &     NULL     &     NULL     &     0     &     0     &    -4.80     &    -0.21     &    18     &         NULL     &         NULL    \\
   56     &    -4.72     &    -4.72     &     1     &     1     &    -4.22     &     0.04     &    37     &         0.32     &         0.32    \\
   60     &     NULL     &     NULL     &     0     &     0     &    -4.76     &     0.78     &    31     &         NULL     &         NULL    \\
  198     &    -3.31     &    -3.31     &     1     &     1     &    -3.14     &    -0.81     &    11     &         0.67     &         0.67    \\
  199     &     NULL     &     NULL     &     0     &     0     &    -6.10     &    -0.97     &     8     &         NULL     &         NULL    \\
  203     &     NULL     &     NULL     &     0     &     0     &    -5.57     &    -1.94     &     2     &         NULL     &         NULL    \\
  204     &     NULL     &     NULL     &     0     &     0     &    -3.39     &    -0.25     &    18     &         NULL     &         NULL    \\
\end{longtable}
% \end{landscape}
%  \label{Tab:isolparam}
}% End onllongtabL

% \end{appendix}

\end{document}